\pdfoutput=1
\documentclass{article}
\newif\iftechreport
\techreporttrue
\usepackage[T1]{fontenc}
\usepackage[utf8]{inputenc}
\usepackage{microtype}
\usepackage{amsmath,amssymb}
\usepackage{graphicx,booktabs,tabularx,array}
\usepackage{multirow,colortbl}
\usepackage{xcolor}
\usepackage{hyperref}
\usepackage{url}
\usepackage{xurl}
\iftechreport
  \usepackage[preprint]{icml2026}
\else
  \usepackage{icml2026}
\fi
\hypersetup{pdftitle={SheetSage2: Coherent Lead-Sheet Transcription with Synthetic Supervision}}
\newcommand{\dsharp}{\raisebox{0.05ex}{\scalebox{0.9}{\textsf{\textbf{x}}}}}

\newcommand{\versiontext}[2]{\iftechreport#2\else#1\fi}
\input{frontmatter.tex}
\iftechreport
  \icmltitlerunning{SheetSage2 Technical Report}
  \hypersetup{pdfauthor={Junyan Jiang, Ruibin Yuan, Jiahao Pan, Wei Xue, Yike Guo, Gus Xia, Yann LeCun}}
\else
  \icmltitlerunning{SheetSage2: Coherent Lead-Sheet Transcription with Synthetic Supervision}
\fi

\begin{document}

\twocolumn[
  \icmltitle{SheetSage2: Coherent Lead-Sheet Transcription with Synthetic Supervision}
  \iftechreport
    \techreportauthors
  \else
    \submissionauthors
  \fi
  \icmlkeywords{Music Transcription, Lead Sheets, Synthetic Data, Structured Decoding, Distillation}
  \vskip 0.3in
]
\iftechreport
  \techreportauthornotes
\else
  \printAffiliationsAndNotice{\icmlEqualContribution}
\fi

\begin{abstract}
Transcribing music into a human-readable score requires a coherent understanding of rhythm, harmony, melody, and form.
Two obstacles limit this goal: annotated recordings are scarce, and accurate local predictions can still produce inconsistent musical sequences.
We present SheetSage2, a unified music transcription framework that combines synthetic data, task-specific structured decoding, and autoregressive distillation.
Automatically annotated MIDI, rendered into audio, provides scalable supervision across music understanding tasks.
Task-specific structured decoders integrate complementary musical cues and their temporal dependencies to produce musically coherent scores.
Autoregressive distillation further retains transcription accuracy without task-specific dynamic programming at inference.
Across eight benchmark collections, a single SheetSage2-AR model exceeds the listed prior systems on 12 of 15 benchmark--metric pairs in our evaluation, substantially improving over SheetSage1 and surpassing task-specific models on several benchmarks.
Model weights and inference code are publicly available\versiontext{}{ at \url{https://huggingface.co/m-a-p/SheetSage2}}.
\end{abstract}

\section{Introduction}
\label{sec:introduction}

A lead sheet represents how a piece of music is organized across multiple musical dimensions.
It describes rhythm through note positions and durations, bar lines, meter, and tempo; harmony through chord symbols and key signatures; and melody through a sequence of pitched notes, with section labels and repeat signs further indicating musical form.
Lead-sheet transcription recovers this representation from audio.
This is a challenging task: a system must infer these complementary dimensions from a polyphonic recording and organize them into a coherent, human-readable score.

Conventional music information retrieval (MIR) systems typically address individual tasks, such as chord estimation~\citep{akram2025chordformer}, key estimation~\citep{kongskey2025}, melody transcription~\citep{donahue2022sheetsage}, and beat tracking~\citep{foscarin2024beatthis}.
However, lead-sheet transcription requires coordination across tasks.
Pitch spelling depends on local key and harmonic context, while note positions and durations must be organized on a shared beat and downbeat grid.
Simply aggregating independent predictions does not ensure this compatibility.

These dependencies motivate treating transcription as a unified sequence-to-sequence task, allowing autoregressive models to learn relationships among musical events.
Recent examples include classical ABC score transcription in TUTTI~\citep{hu2026tutti} and multi-instrument MIDI transcription in MuScriptor~\citep{rouard2026muscriptor}.

However, supervised training of such models depends on large collections of complete, well-aligned audio--score pairs, which remain scarce.
Existing MIR datasets often annotate only selected attributes, such as beats, chords, keys, or melody, rather than complete lead sheets.
Even available labels can have imprecise timing~\citep{donahue2022sheetsage}.
These limitations make heterogeneous MIR annotations difficult to use directly as targets for unified score generation.

We address this data bottleneck with SheetSage2, combining local prediction with autoregressive sequence modeling.
First, \textbf{SheetSage2-Prober} learns frame-level prediction using task-specific objectives that accommodate partial annotations and varying label quality.
Its structured decoders produce complete event sequences while encouraging consistent tempo, meter, and melodic register.
We then train \textbf{SheetSage2-AR} on these sequences through \emph{autoregressive distillation}.
The student improves several benchmark metrics while largely retaining the Prober's transcription consistency, without task-specific dynamic programming at inference.

To further expand supervision, we synthesize audio from large MIDI collections and derive aligned task labels through symbolic analysis and pseudo-label bootstrapping.
These examples complement partially annotated real recordings in Prober training.

Our contributions are threefold:
\begin{enumerate}
    \item \textbf{MIDI-derived supervision for audio music understanding.}
    We extend content-based MIDI self-annotation into a scalable synthetic-audio supervision pipeline and demonstrate strong performance across multiple music-understanding benchmarks on real recordings.
    \item \textbf{Structured decoding for unified transcription.}
    We develop task-specific decoders with explicit tempo, meter, and pitch context, improving transcription consistency and coherence within and among tasks.
    \item \textbf{Autoregressive distillation.}
    We transfer the structured teacher outputs to an AR student that dispenses with task-specific dynamic programming and maintains strong performance across the evaluated tasks.
\end{enumerate}

\versiontext{}{YuE2~\citep{yuan2026yue2} first introduced SheetSage2 and the MERT2 encoder to supply symbolic and semantic supervision for music generation; this report describes SheetSage2's method and evaluation in full.}

\section{Methodology}
\label{sec:methodology}

\subsection{SheetSage2-AR}
\label{sec:sheetsage2-ar}

\begin{figure*}[t]
    \centering
    \includegraphics[width=\textwidth]{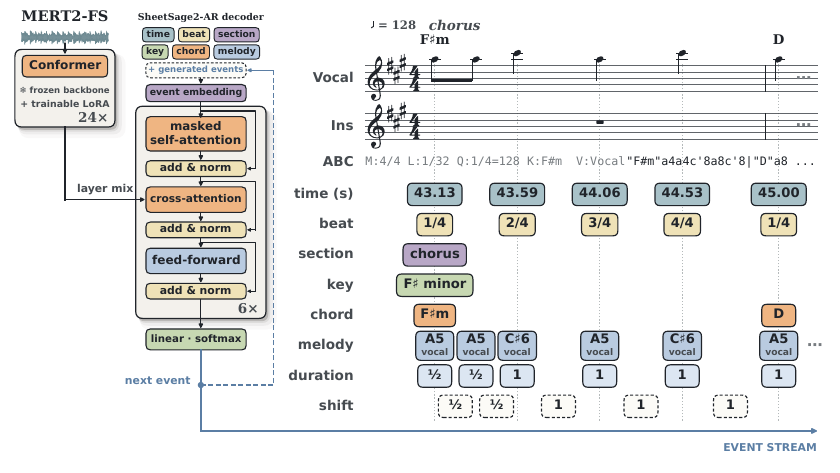}
    \caption{\textbf{SheetSage2-AR inference.} Audio representations and preceding events condition a unified event sequence, which is converted into ABC notation and an editable lead sheet. The right panel aligns an illustrative score, its ABC notation, and its events. The excerpt is serialized independently, with the active key emitted at its first beat and omitted thereafter unless it changes. Musical units are used for readability.}
    \label{fig:sheetsage2-ar}
\end{figure*}

We first describe SheetSage2-AR, the final model produced by our training pipeline. It transcribes audio into a single event sequence representing rhythm, harmony, melody, and musical form (Figure~\ref{fig:sheetsage2-ar}).
A \versiontext{MERT-v2-FullSong}{\href{https://huggingface.co/m-a-p/MERT-v2-FullSong}{MERT-v2-FullSong}} (MERT2-FS) encoder~\citep{yuan2026yue2} from the MERT family~\citep{li2023mert,yuan2023marble} processes mono audio $x$ at 24~kHz. An autoregressive decoder attends to the audio representations and preceding tokens $y_{<n}$ to predict $p_\theta(y_n\mid y_{<n},x)$. At inference, greedy decoding selects the highest-probability token allowed by the event grammar.

The event language combines absolute beat timestamps with relative positions on a sixteenth-note grid. Typed tokens encode meter, beat position, section, key, chord, melody pitch and source, and note duration. Timestamps are quantized into 10-ms bins. A deterministic notation builder converts the decoded sequence into an ABC score or a MIDI file. Appendix~\ref{app:ar-format} specifies the event format and decoding constraints with token-level examples, and Appendix~\ref{app:full-song} shows lead-sheet excerpts.

\subsection{An Overview of the Training Pipeline}
\label{sec:data-pipeline}

Our training pipeline has three stages.

1. \textbf{Synthetic data creation.}
Training on the small set of real recordings we acquired yields limited performance. We derive task labels from symbolic MIDI files and render the same performances as audio, producing aligned supervision at scale.

2. \textbf{Prober training.}
Mixed data with inaccurate timestamps and incomplete or inconsistent annotations cannot provide complete, consistent event-sequence targets for SheetSage2-AR. We therefore first train \textbf{SheetSage2-Prober}, two frame-wise prober models that learn from partial and noisy annotations. Structured decoding converts their predictions into discrete, musically consistent labels.

3. \textbf{Autoregressive distillation.} We train SheetSage2-AR only on pseudo-labels generated by SheetSage2-Prober. The teacher's structured decoders produce complete event sequences, which serve as hard targets for the AR student to learn the unified lead-sheet representation in a sequence-to-sequence manner.

\subsection{Synthetic Music \& Synthetic Labels}
\label{sec:midi-labels}

\begin{figure*}[t]
    \centering
    \includegraphics[width=\textwidth]{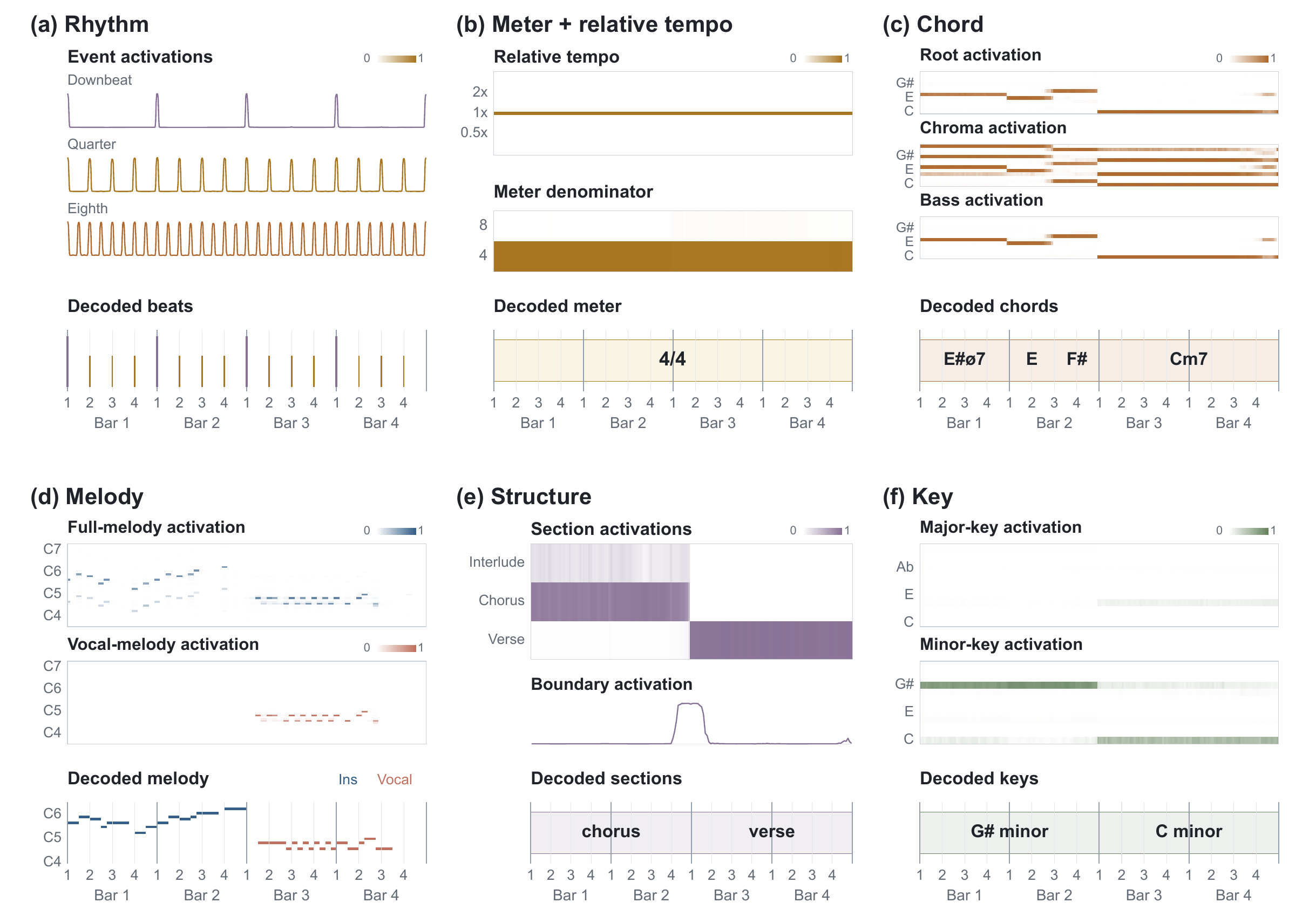}
    \caption{\textbf{Real Prober activations and structured decoding on RWC-MDB-P-2001 No.~1.}
    (a) Rhythm, (b) meter and relative tempo, (c) chord, (d) melody, (e) selected section and boundary activations, and (f) key.
    Thin grid lines indicate beats and bold lines downbeats.}
    \label{fig:prober-example}
\end{figure*}

MIDI files are a common format for representing music symbolically. MIDI collections support symbolic music generation and analysis by representing note pitches, durations, instruments, and timing~\citep{raffel2016extracting,jiang2025versatile}.
Rendering these scores also provides aligned audio--note pairs and instrument stems for transcription and source separation, as in Slakh~\citep{manilow2019cutting}.
However, higher-level labels in MIDI files, such as key, chords, and melody tracks, are often missing or unreliable.
For Beatles MIDI files, \citet{raffel2016extracting} reported a MIREX weighted key score of 0.400, rising to 0.842 after excluding C-major metadata, still below curated annotations.
Curated chord and melody resources remain small; RWC-Pop, for example, covers 100 songs~\citep{goto2006aist,cho2011feature}.

MIDI without curated annotations is far more abundant: the Los Angeles MIDI Dataset contains 404,714 deduplicated files~\citep{lev2024losangelesmididataset}.
Its explicit notes and instrument tracks facilitate symbolic analysis without first estimating them from audio~\citep{raffel2016extracting}.
We therefore label MIDI with symbolic MIR models and transfer this supervision to rendered audio.

\paragraph{Pseudo-label bootstrapping.}
MIDI-based music analysis is generally more reliable than audio-based analysis, as MIDI provides explicit multi-track symbolic information.
To create high-quality melody, chord, and key labels for MIDI files, we first use task-specific expert rules or metadata to curate a small seed set with high precision. We then fine-tune pretrained symbolic models~\citep{jiang2025versatile} on these seed labels to obtain a \textit{MIDI auto-labeler}.
For example, the melody seed pool contains 17,134 MIDI files with identifiable melody-track names, and bootstrapping expands it to 185,887 annotated songs (10.8$\times$).
Table~\ref{tab:synthetic-labels} lists the label sources; Appendix~\ref{app:pseudolabel-bootstrapping} describes the bootstrapping procedure.

\begin{table}[t]
\centering\small
\caption{Synthetic labels by task.}
\label{tab:synthetic-labels}
\begin{tabularx}{\columnwidth}{@{}lX@{}}
\toprule
Task & Label source \\
\midrule
Beat / tempo & Tick resolution and tempo events. \\
Meter / downbeat & Time-signature events and the metrical grid. \\
Melody & Bootstrapped from track-name labels. \\
Chord & Bootstrapped from expert-rule labels. \\
Key & Bootstrapped from expert-rule labels. \\
Structure & Curated MIDI text markers (SLMS)~\citep{eldeeb2025barwise}. \\
\bottomrule
\end{tabularx}
\end{table}

The annotated Los Angeles collection~\citep{lev2024losangelesmididataset,mald2025dataset} and SLMS supply approximately 361,000 synthetic source records (Appendix~\ref{app:training}).
We render MIDI with FluidSynth and the FluidR3 GM SoundFont, pairing the labels with 24-kHz audio for Prober training. Section~\ref{sec:synthetic-data} shows that even this simple rendering provides useful supervision, improving most task averages on real recordings.

\subsection{SheetSage2-Prober}
\label{sec:prober}

Both human and synthetic annotations contain errors, biases, and missing attributes.
Tapped beat annotations can deviate by tens of milliseconds~\citep{driedger2019correcting}, exceeding our 10-ms output resolution.
Musical ambiguity also permits alternative tempo levels~\citep{schreiber2020tempo} and chord interpretations~\citep{koops2020personalization}.
Rather than treating any single annotation as exact, the Prober learns local probabilities from the available labels, masks missing targets, and uses source hints to accommodate different annotation conventions. 
Instead of using autoregressive targets, the Prober produces frame-wise outputs, making the loss less sensitive to small timing errors in the ground-truth annotations.

SheetSage2-Prober comprises two separately trained acoustic models, each with its own adapted MERT2-FS encoder.
The non-melody model predicts 160 binary features at 25~Hz; four slots per rhythm event provide 100-Hz timing.
The melody model operates on subdivisions of the predicted beat grid, conditioned on a vocal or full-melody hint.
The non-melody model uses a two-layer MLP on the final encoder states; the melody model uses learned layer mixing and beat-aligned Transformer layers.
Figure~\ref{fig:prober-example} illustrates their outputs, with probabilities denoted as follows:

\begin{itemize}
    \item \textbf{Rhythm events} ($b_4^{(i)},b_8^{(i)},d^{(i)}\in[0,1]$): quarter-note, eighth-note, and downbeat probabilities on the 100-Hz grid.
    \item \textbf{Meter denominator} ($\eta_4^{(i)},\eta_8^{(i)}\in[0,1]$): support for denominators 4 and 8; the numerator is inferred during decoding.
    \item \textbf{Median tempo} ($g^{(i)}\in[0,1]^{25}$): support for quarter-note tempo bins derived from the excerpt's median pulse interval.
    \item \textbf{Relative tempo} ($r^{(i)}\in[0,1]^{25}$): support for bins of the local-to-median tempo ratio.
    \item \textbf{Chord} ($c^{(i)}\in[0,1]^{12\times4}$): pitch-class probabilities for root, chroma, extension, and bass.
    \item \textbf{Key} ($k^{(i)}\in[0,1]^{24}$): support for 12 major and 12 minor keys.
    \item \textbf{Section boundary} ($s_o^{(i)}\in[0,1]$): probability of a boundary within $\pm7$ frames at 25~Hz.
    \item \textbf{Section label} ($s^{(i)}\in[0,1]^{23}$): support for 23 section classes.
    \item \textbf{Markovian melody} ($\mu^{(i)}\in[0,1]^{926}$): one categorical distribution on the sixteenth-note grid: $128\times7$ pitch--interval categories, 24 durations, one rest, and five auxiliary slots (Section~\ref{sec:melody-decoding}). A global hint embedding selects whether the output covers only the vocal melody or the full melody.
\end{itemize}

Non-melody features use binary cross-entropy losses; melody uses categorical cross-entropy, excluding missing targets.
Appendix~\ref{app:prober-outputs} details the channels and tempo bins.

\subsection{Structured Decoding}
\label{sec:structured-decoding}

MIR pipelines commonly decode frame-wise predictions with probabilistic sequence models such as hidden Markov models (HMMs) and conditional random fields (CRFs)~\citep{bock2016joint,jiang2019large}.
SheetSage2-Prober follows this approach with a six-stage decoder.
Each stage fixes a decision that constrains subsequent dependent stages: local tempo is interpreted relative to one shared anchor, and later attributes are placed on the decoded rhythm grid.
This encourages long-term tempo coherence and ensures that the decoded attributes share a compatible musical grid.

The decoding steps are as follows:

1. \textbf{Tempo anchor selection.} We pool the median-tempo predictions $g^{(i)}$ to select a single anchor $\hat g$ as the global tempo reference (Section~\ref{sec:beat-decoding}).

2. \textbf{Joint decoding of pulses and local tempo.} Conditioned on the tempo anchor, we jointly decode eighth-note pulse positions and their intervals, which determine the local tempo (Section~\ref{sec:beat-decoding}).

3. \textbf{Joint decoding of downbeats and meter.} Conditioned on the fixed pulse sequence, we decode metrical positions and meter changes, identifying the beats and downbeats (Section~\ref{sec:beat-decoding}).

4. \textbf{Key and structure decoding.} Conditioned on the rhythm grid, we decode keys and section labels at the measure level, restricting their changes to downbeats (Appendix~\ref{app:harmony-decoding}).

5. \textbf{Chord decoding.} Conditioned on the beat grid, we decode chords with boundaries aligned to beats. The local key then determines their enharmonic spelling (Appendices~\ref{app:harmony-decoding} and~\ref{app:spelling}).

6. \textbf{Melody decoding.} Conditioned on subdivisions of the beat grid, we decode a melody sequence from the Markovian predictions $\mu^{(i)}$ (Section~\ref{sec:melody-decoding}).

\subsection{Structured Decoding: Rhythm}
\label{sec:beat-decoding}

Long-term rhythm inconsistency~\citep{fuentes2019structure}, often manifested as switching between half- and double-tempo interpretations~\citep{foscarin2026maskeddiffusion}, is a common issue in beat tracking and is particularly harmful for lead-sheet transcription.
Consider a song with two verses that share similar rhythmic content, one without drums and one with drums.
Local predictions may favor 80~BPM in the first verse and 160~BPM in the second, even though the underlying musical tempo has not changed.
Such a switch gives corresponding passages different beat grids and notated durations.
Instead of a dynamic Bayesian network (DBN) beat tracker, which models tempo only locally~\citep{bock2016joint}, we design a three-stage rhythm decoder that encourages global tempo consistency.

First, we pool the median-tempo activations $g^{(i)}$ across the recording and select the tempo bin in 60--240~BPM with the highest mean activation as the anchor $\hat g$.
In the example above, the predictions may support bins near both 80 and 160~BPM, but pooling selects one reference for both verses.
The anchor stays fixed, while relative-tempo activations $r^{(i)}$ let local tempo vary around it.

Second, we recover the eighth-note rhythm grid.
Pulse positions lie on the 100-Hz grid (10-ms precision).
At position $i$, an outgoing pulse interval $\delta$ implies quarter-note tempo $\operatorname{BPM}(\delta)=3000/\delta$.
Let $F^{(i)}(\delta)$ be the best path score for a pulse at $i$ with this outgoing interval.
The pulse dynamic program is
\begin{equation}
\begin{split}
F^{(i)}(\delta)={}&h(b_8^{(i)})+W^{(i)}(\delta)\\
&+\max_{\delta'\in\mathcal D,\,\delta'\leq i}
\Bigl[F^{(i-\delta')}(\delta')-10\log^2\!\frac{\delta'}{\delta}\Bigr],
\end{split}
\label{eq:pulse-dp}
\end{equation}
where $W^{(i)}(\delta)=W\bigl(\hat g,\operatorname{BPM}(\delta);r^{(i:i+\delta)}\bigr)$ scores how well the local activations $r^{(i:i+\delta)}$ over the candidate interval support tempo $\operatorname{BPM}(\delta)$ given anchor $\hat g$.
Higher scores indicate stronger support.
Here $h(v)=15(v-0.15)$, and $\mathcal D=\{12,\ldots,55\}$ covers about 54.5--250~BPM.
The log-ratio penalty discourages abrupt changes between adjacent pulse intervals, while $W$ evaluates local tempo against the shared global reference.
Appendix~\ref{app:tempo-evidence} gives the tempo evidence score and pulse-path boundary conditions.

Finally, conditioned on the recovered pulses, a second dynamic program jointly decodes metrical positions and meter changes to construct the beat grid.
It follows the bar-position model of the joint beat and downbeat tracker of \citet{bock2016joint}, but additionally allows local meter changes, such as incomplete bars, and global meter changes.
Appendix~\ref{app:decoders} gives the recurrences and hyperparameters.

\subsection{Structured Decoding: Markovian Melody}
\label{sec:melody-decoding}

Melody decoding faces another difficulty: octave ambiguity.
The Hooktheory annotations used by SheetSage specify relative octaves without a reliable absolute register~\citep{donahue2022sheetsage}, while melody extractors can confuse octave-related pitches~\citep{salamon2012melody,chen2022tonet}.
The faint activations one octave below the decoded melody in Figure~\ref{fig:prober-example}(d) illustrate this ambiguity.

In SheetSage2, we focus on suppressing local octave errors, such as transcribing C4--E4--G4 as C5--E4--G4, which distort the intervals within a melody. 
Even when the absolute octave is ambiguous, intervals between consecutive pitches remain informative and provide more stable supervision.
To achieve this, instead of directly predicting a MIDI pitch at each onset, SheetSage2-Prober assigns a separate score to each pitch--interval category, making the onset score depend on the preceding pitch.
In the previous example, the score assigned to E4 depends on the preceding pitch in the candidate melody (C4 or C5).

At position $i$, $\mu^{(i)}_{p,b}$ denotes the probability of an onset with MIDI pitch $p\in\{0,\ldots,127\}$ and interval bucket $b\in\{0,\ldots,6\}$.
The seven buckets group signed semitone intervals from the preceding onset, with ranges $(-\infty,-18]$, $[-17,-12]$, $[-11,-6]$, $[-5,5]$, $[6,11]$, $[12,17]$, and $[18,\infty)$ in order $b=0,\ldots,6$.
For a candidate onset $p$ and a preceding onset $p'$ less than 32 grid positions earlier, the transition score is $ s_{\mathrm{on}}^{(i)}(p'\!\to p)=\log\bar\mu^{(i)}_{p,\,B(p-p')} $
where $B$ maps the pitch interval to its bucket and $\bar\mu$ denotes the adjusted decoding probabilities.

In the example above, C4--E4 and C5--E4 imply intervals of $+4$ and $-8$ semitones and therefore use buckets $b=3$ and $b=2$, respectively, to score E4.
These onset scores then guide dynamic programming over candidate melody sequences.
Appendix~\ref{app:melody-dp} specifies the probability adjustments, the token grammar, and the recurrences.

\section{Experiments}
\label{sec:experiments}

\subsection{Main Results}
\label{sec:main-results}

We evaluate transcription performance on rhythm, key, chords, structure, and melody across eight benchmark collections: GTZAN, osu2017, GiantSteps, Chords1217, JAAH, HarmonixSet, RWC-Pop, and Rock Corpus~\citep{tzanetakis2002musical,jiang2026osu2017,knees2015two,humphrey2015four,eremenko2018jaah,nieto2019harmonix,goto2002rwc,temperley2013rock}.
Appendix~\ref{app:evaluation} gives the protocols, baseline provenance, and supplementary results.

\begin{table*}[!t]
\centering
\caption{\textbf{Multitask transcription on real recordings.} SheetSage2-Prober and SheetSage2-AR are compared with SheetSage1, madmom~\citep{madmom}, and prior task-specific systems. Scores are percentages; higher is better. Bold and underlining mark the largest and second-largest point estimates in each benchmark--metric pair.}
\label{tab:main-results}
\begingroup
\small
\setlength{\tabcolsep}{3pt}
\renewcommand{\arraystretch}{1.02}
\begin{tabular*}{\textwidth}{@{\extracolsep{\fill}}cccccccc@{}}
\toprule
\multirow{2}{*}{\textbf{Task}} & \multirow{2}{*}{\textbf{Benchmark}} &
\multirow{2}{*}{\textbf{Metric}} & \multirow{2}{*}{\textbf{SheetSage1}} &
\multirow{2}{*}{\textbf{madmom}} &
\multirow{2}{*}{\shortstack{\textbf{Prior}\\\textbf{specialist}}} &
\multicolumn{2}{c}{\textbf{SheetSage2}} \\
 & & & & & & \textbf{Prober} & \textbf{AR} \\
\midrule
\multirow{2}{*}{\textbf{Beat}} & GTZAN & \multirow{2}{*}{F1\,$\uparrow$} & 86.07 & 86.07 & \textbf{89.01}$^{a}$ & 82.93 & \underline{86.27} \\
 & osu2017 & & 91.80 & 91.80 & 89.19$^{a}$ & \underline{92.28} & \textbf{93.01} \\
\cmidrule(lr){1-8}
\multirow{2}{*}{\textbf{Downbeat}} & GTZAN & \multirow{2}{*}{F1\,$\uparrow$} & 64.65 & 64.65 & 78.28$^{a}$ & \underline{78.74} & \textbf{80.45} \\
 & osu2017 & & 83.47 & 83.47 & 85.90$^{a}$ & \underline{92.79} & \textbf{92.90} \\
\cmidrule(lr){1-8}
\multirow{2}{*}{\textbf{Key}} & GiantSteps & \multirow{2}{*}{Score\,$\uparrow$} & 43.89 & 74.62 & 72.09$^{b}$ & \textbf{78.29} & \underline{77.73} \\
 & GTZAN & & 54.56 & 72.05 & \underline{74.43}$^{b}$ & 72.62 & \textbf{75.77} \\
\cmidrule(lr){1-8}
\multirow{3}{*}{\textbf{Chord}} & osu2017 & \multirow{3}{*}{Maj/min\,$\uparrow$} & 79.82 & 77.42 & 86.55$^{c}$ & \textbf{90.43} & \underline{90.08} \\
 & Chords1217 & & 72.98 & 83.52$^{*}$ & \underline{83.94}$^{c,\dagger}$ & \textbf{84.29} & 83.81 \\
 & JAAH & & 39.77 & 51.24 & 59.45$^{c}$ & \underline{62.94} & \textbf{64.50} \\
\cmidrule(lr){1-8}
\multirow{3}{*}{\textbf{Structure}} & \multirow{3}{*}{HarmonixSet} & Accuracy\,$\uparrow$ & --- & --- & 80.03$^{d}$ & \textbf{80.78} & \underline{80.51} \\
 & & F1 (0.5\,s)\,$\uparrow$ & --- & --- & \textbf{70.63}$^{d}$ & 66.95 & \underline{67.96} \\
 & & F1 (3\,s)\,$\uparrow$ & --- & --- & 79.50$^{d}$ & \underline{81.65} & \textbf{82.86} \\
\cmidrule(lr){1-8}
\multirow{3}{*}{\textbf{Melody}} & \multirow{2}{*}{RWC-Pop} & Vocal F1\,$\uparrow$ & 62.71 & --- & 62.71$^{e}$ & \textbf{83.08} & \underline{82.51} \\
 & & Full F1\,$\uparrow$ & 64.02 & --- & 64.02$^{e}$ & \underline{75.00} & \textbf{75.29} \\
 & Rock Corpus & Vocal F1\,$\uparrow$ & 49.19 & --- & 49.19$^{e}$ & \underline{65.98} & \textbf{67.08} \\
\bottomrule
\end{tabular*}
\endgroup
\par\smallskip
{\footnotesize\raggedright
$^{a}$Beat This!~\citep{foscarin2024beatthis}, averaging three seeds;
$^{b}$S-KEY~\citep{kongskey2025};
$^{c}$ChordFormer~\citep{akram2025chordformer};
$^{d}$SongFormer~\citep{hao2025songformer};
$^{e}$SheetSage1~\citep{donahue2022sheetsage}.
SheetSage1 reuses madmom rhythm; $^{e}$ repeats its melody scores.
$^{*}$Madmom's chord training set overlaps with Chords1217. $^{\dagger}$ChordFormer performs 5-fold cross-validation on Chords1217.
Melody F1 ignores offsets and octaves. See Appendix~\ref{app:evaluation}.\par}
\end{table*}

SheetSage2-AR exceeds the strongest listed prior baseline on 12 of the 15 benchmark--metric pairs.
SheetSage2-AR exceeds SheetSage2-Prober on 10 of the 15 benchmark--metric pairs, including both beat and downbeat benchmarks. This shows that autoregressive distillation largely preserves Prober performance with a single event-generation interface.

These accuracy metrics do not fully capture transcription consistency: pitch-class melody F1 ignores octave assignment, while beat F1 does not directly measure long-term tempo consistency. Sections~\ref{sec:tempo-consistency} and~\ref{sec:octave-consistency} therefore complement these scores with tempo- and octave-consistency analyses.

\subsection{Effectiveness of Synthetic Data}
\label{sec:synthetic-data}

To assess the contribution of synthetic data, we compare SheetSage2-Prober trained on real-only, synthetic-only, and combined task data.
Table~\ref{tab:synthetic-data} summarizes the 15 benchmark--metric pairs in Table~\ref{tab:main-results}; Appendix~\ref{app:synthetic-data} gives every score.

Adding synthetic data improves the beat, downbeat, chord, and structure task averages over both single-source conditions, while leaving the melody average nearly unchanged and slightly reducing the key average.
The synthetic-only model also remains competitive on chord and key estimation and reaches 57.76\% vocal-melody F1 on RWC-Pop, even though its synthetic training audio contains no singing.
This suggests that training on acoustically simplified audio can transfer to real recordings, including vocal melody.
Appendix~\ref{app:synthetic-data} shows this on one RWC-Pop opening phrase (Figure~\ref{fig:synth-melody-example}). Together, these results support using synthesized music to enhance understanding on real recordings.

\begin{table*}[!t]
\centering
\caption{\textbf{Effect of task-training data.} Each entry is the unweighted mean (\%) of a task's benchmark--metric pairs in Table~\ref{tab:main-results}. Structure averages accuracy and both boundary F1 scores; melody averages RWC-Pop vocal/full F1 and Rock Corpus vocal F1. Bold marks the best task average.}
\label{tab:synthetic-data}
\small
\setlength{\tabcolsep}{4pt}
\begin{tabular*}{\textwidth}{@{\extracolsep{\fill}}lrrrrrr@{}}
\toprule
Task-training audio & Beat & Downbeat & Key & Chord & Structure & Melody \\
\midrule
Synthetic only & 79.52 & 76.64 & 75.66 & 76.99 & 56.76 & 54.77 \\
Real only & 85.43 & 82.06 & \textbf{75.69} & 77.65 & 74.88 & 74.68 \\
Real + synthetic & \textbf{87.60} & \textbf{85.77} & 75.46 & \textbf{79.22} & \textbf{76.46} & \textbf{74.69} \\
\midrule
Improvement by synth. & +2.17 & +3.71 & -0.24 & +1.57 & +1.58 & +0.01 \\
\bottomrule
\end{tabular*}
\end{table*}

\begin{figure*}[!t]
\centering
\includegraphics[width=\textwidth]{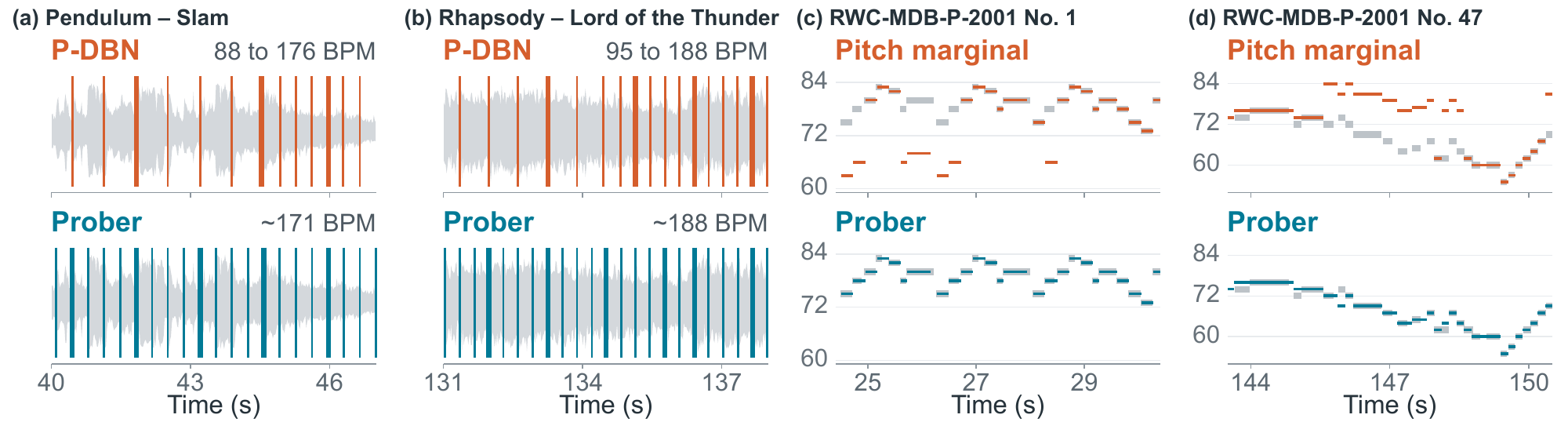}
\caption{\textbf{Selected examples of structured decoding.}
(a,b) P-DBN (orange) changes tempo level over several beats, while Prober (blue) remains stable. Lines mark decoded beats and thick lines denote downbeats.
(c,d) Pitch-marginal decoding introduces octave shifts that Prober avoids. Gray bars show reference notes.}
\label{fig:structured-decoding-examples}
\end{figure*}

\subsection{Effect of Structured Decoding on Rhythmic Consistency}
\label{sec:consistency}
\label{sec:tempo-consistency}

We evaluate the effectiveness of structured decoding in improving global tempo consistency.
The proposed rhythm decoder has three stages: median-tempo anchor selection, joint pulse/local-tempo decoding, and beat-grid decoding.
We compare it with a conventional DBN decoder~\citep{bock2016joint} applied to the same Prober predictions (P-DBN).
This baseline uses only quarter-beat and downbeat activations to jointly infer tempo and metrical positions, without the predicted tempo anchor.

We measure the \emph{mixing ratio} on osu2017 and Hooktheory: the percentage of recordings whose predictions mix metrical levels relative to the reference, such as both $1\times$ and $2\times$ tempo within one recording.
A higher ratio indicates less consistent metrical interpretation, which can complicate score reading.
Appendix~\ref{app:consistency} provides the scoring details and distributions.

Table~\ref{tab:tempo-consistency} shows that Prober and AR mix tempo levels much less often than Beat This! on both datasets.
The proposed decoder also shows better consistency compared to P-DBN. 
Figure~\ref{fig:structured-decoding-examples}(a,b) shows two Hooktheory excerpts where P-DBN changes tempo level while Prober keeps a stable grid.

\begin{table*}[!t]
\centering
\caption{\textbf{Global tempo consistency.} Mixing ratio (\%) on 135 common osu2017 and 2,943 common Hooktheory recordings; lower is better. P-DBN applies madmom's DBN to Prober activations.}
\label{tab:tempo-consistency}
\small
\setlength{\tabcolsep}{2pt}
\begin{tabular*}{\textwidth}{@{\extracolsep{\fill}}l*{8}{c}@{}}
\toprule
 & \multicolumn{4}{c}{\textbf{osu2017}} & \multicolumn{4}{c}{\textbf{Hooktheory}} \\
\cmidrule(lr){2-5}\cmidrule(lr){6-9}
 & Beat This! & Prober & P-DBN & AR & Beat This! & Prober & P-DBN & AR \\
\midrule
Mixing ratio (\%) & 20.74 & 2.96 & 3.70 & \textbf{1.48} & 17.30 & \textbf{0.34} & 1.36 & 0.51 \\
\bottomrule
\end{tabular*}
\end{table*}

\subsection{Effect of Structured Decoding on Octave Consistency}
\label{sec:octave-consistency}

We next test whether the melody decoder's local pitch-interval dependencies improve octave consistency.
We compare the full Prober decoder with a pitch-marginal variant that removes dependence on the previous pitch while retaining the trained model and the remaining decoding rules.
On RWC full melody, the \emph{mixing ratio} is the percentage of contiguous instrument passages that mix octaves.

On 333 common passages, full decoding reduces the mixing ratio from 12.91\% to 9.31\% (Table~\ref{tab:octave-consistency}).
SheetSage1 mixes octaves in 34.83\% of passages, while AR reaches 11.71\% without an explicit melody DP at inference.
These results show that local pitch dependencies can improve consistency over a melodic passage.
Figure~\ref{fig:structured-decoding-examples}(c,d) shows two examples where the full decoder avoids octave shifts introduced by pitch marginalization.
Appendix~\ref{app:consistency} reports more detailed results on octave consistency.

\begin{table}[htbp]
\centering
\caption{\textbf{Octave consistency on RWC full melody.} Mixing ratio (\%) on 333 contiguous instrument passages common to all systems; lower is better.}
\label{tab:octave-consistency}
\small
\setlength{\tabcolsep}{2pt}
\begin{tabular*}{\columnwidth}{@{\extracolsep{\fill}}lcccc@{}}
\toprule
 & SheetSage1 & Prober & \shortstack{Prober\\(Pitch Marginal)} & AR \\
\midrule
Mixing ratio (\%) & 34.83 & \textbf{9.31} & 12.91 & 11.71 \\
\bottomrule
\end{tabular*}
\end{table}

\section{Related Work}
\label{sec:related-work}

\paragraph{Automatic Music Transcription.}
Piano transcription systems recover precise note and pedal events~\citep{kong2021highresolution}, but do not directly produce scores.
SheetSage combines beat tracking, key, chord, and melody estimation for lead-sheet transcription~\citep{donahue2022sheetsage}.
Unified models include TUTTI for score transcription~\citep{hu2026tutti}, and MT3 and MuScriptor for multi-instrument event transcription~\citep{gardner2022mt3,rouard2026muscriptor}.
Generalization across diverse recordings remains limited by the availability of aligned annotations~\citep{rouard2026muscriptor}.

\paragraph{Synthetic Supervision and Label Mining.}
Large MIDI collections support scalable synthetic supervision~\citep{raffel2016learning,lev2024losangelesmididataset}.
Synthesized audio has been used for transcription~\citep{hu2026tutti,rouard2026muscriptor} and generative music editing~\citep{zhang2025instructmusicgen}.
Label mining extends MIDI supervision to structure analysis through SLMS~\citep{eldeeb2025barwise}, while content-based self-annotation provides chord and key labels~\citep{jiang2025musiclabelsitself}.

\paragraph{Structured Decoding.}
MIR systems often decode frame-wise activations with DBNs for beats, downbeats, and tempo~\citep{bock2016joint} and CRFs for chords~\citep{jiang2019large}; skip-chain CRFs further link distant passages at the cost of loopy inference~\citep{fuentes2019structure}.
Recent beat trackers drop DBNs~\citep{foscarin2024beatthis} or use masked diffusion for coherent beat grids~\citep{foscarin2026maskeddiffusion}.
Metrical-level and octave errors persist~\citep{schreiber2020tempo,salamon2012melody,chen2022tonet}; we counter them with a shared tempo anchor and interval-dependent melody scores.

\paragraph{Sequence-Level Distillation.}
Our autoregressive training resembles sequence-level distillation~\citep{kim2016sequence}, which can retain translation quality without beam search.
Similarly, our student learns from structured teacher outputs without the teacher's decoders at inference.

\section{Limitations}
\label{sec:limitations}
SheetSage2 makes strong simplifying assumptions that limit its generalization to a wider range of musical styles. \textbf{Rhythmic assumptions}: Notes lie on a sixteenth-note grid, so triplets and finer subdivisions can only be approximated. Decoding fixes one tempo anchor for the whole recording and interprets local tempo relative to it, which suits songs with a single prevailing tempo but not pieces with large tempo changes; the meter inventory is also limited. \textbf{Harmonic assumptions}: The model assumes 12-tone equal temperament. The chord vocabulary covers triads, suspended, sixth, and seventh chords (Table~\ref{tab:chord-vocabulary}), so extended and altered chords such as ninths cannot be represented. \textbf{Limited genre generalizability}: Both the lead-sheet format and our training data target popular music with a steady beat.
SheetSage2 is not designed for classical music, which often requires several voices and expressive tempo, or for free-rhythm music without a metrical grid.

\section{Conclusion}
\label{sec:discussion}

We present SheetSage2-AR, a unified lead-sheet transcription system that uses a MERT2-FS encoder to infer rhythm, harmony, melody, and form from audio.
To address limited annotation coverage, we combine real recordings with synthetic audio and labels derived from large symbolic music collections.
Structured decoding promotes score consistency, and autoregressive distillation transfers these structured outputs to a model that needs no task-specific dynamic programming at inference.

With more training data, some explicit musical constraints may become less necessary.
As annotated recordings remain scarce, expanding synthetic data with high-quality labels is promising, motivating advances in symbolic music analysis and generation for better, more diverse synthetic supervision.

\label{main-text-end}
\section*{AI Use Statement}
In this work, we used generative AI tools (LLM-based coding and writing assistants) to help implement and debug method and experiment code, including inference, evaluation, and analysis scripts.
Additionally, we used generative AI tools to aid and polish writing, create figures, and search for and format references.
The authors reviewed and tested AI-assisted code, and reviewed and edited all AI-assisted text, figures, and references.
We take responsibility for the final content of this work, including text, claims or artifacts produced with the aid of generative AI.

\section*{Impact Statement}
Prober training combines MIDI renderings from public symbolic datasets, licensed Hooktheory annotations, a private lead-sheet dataset, and the HarmonixSet training set.
SheetSage2-AR is trained primarily on CC0 music and synthetic data, with most synthetic training data supplied under license.
The work involves no human-subject study.

Automatic transcription can facilitate copying compositions from recordings; a transcribed lead sheet does not grant permission to reproduce or distribute a work, and users must consider the applicable rights and attribution.
Transcriptions can contain errors and should not be treated as authoritative scores.

The training data and evaluation are dominated by popular music in the Western tonal tradition (Section~\ref{sec:limitations}), and the model likely serves these styles better than others.
Widely used transcription tools that favor such music could further marginalize underrepresented musical traditions, whose tuning, meter, or notation may not fit the lead-sheet format.

\section*{Reproducibility Statement}
Section~\ref{sec:methodology} and Appendix~\ref{app:ar-format} specify the event format and the AR decoding constraints.
Appendix~\ref{app:training} lists the training sets, model configurations, and Prober output channels, and Appendix~\ref{app:pseudolabel-bootstrapping} describes the label bootstrapping.
Appendix~\ref{app:decoders} gives the decoder recurrences and hyperparameters.
Appendix~\ref{app:evaluation} states the evaluation cohorts, exclusions, metrics, and baseline settings, and Appendix~\ref{app:consistency} defines the consistency diagnostics.

\versiontext{Model weights and inference code, including a preset for the paper's inference settings, are publicly available; the link is omitted for anonymous review.}{Model weights and inference code, including a preset for the paper's inference settings and the benchmark scores, are available at \url{https://huggingface.co/m-a-p/SheetSage2}.}
Part of the training data is private or licensed, and some benchmark audio is distributed only under its providers' terms; these descriptions document the experiments without implying redistribution of that audio.

\bibliography{references}
\bibliographystyle{icml2026}
\clearpage
\appendix
\onecolumn
\section{Complete AR Event Format and Inference Details}
\label{app:ar-format}

\paragraph{Task prefix.}
The supplied prefix starts with \texttt{<|sos|>}, lists the requested tasks, and ends with \texttt{<|out|>}. Full transcription requests timestamps, downbeats and meter, structure, key, full-vocabulary chords, and full melody. Generated events follow, ending with \texttt{<|eos|>}.

\paragraph{A unified musical event language.}
All attributes share one vocabulary and one chronological sequence.
Within each event group, the field order is subbeat shift, timestamp, meter and beat position, structure, key, chord, then melody. Fields that need no update are omitted.

\begin{itemize}
    \item \textbf{Absolute beat time.}
    A token such as \texttt{<time\_8.48s>} gives a beat onset relative to the beginning of the current audio window, quantized at 100~Hz (10~ms).
    Successive anchors define the mapping between musical position and audio time, including local tempo variation.

    \item \textbf{Beat position and meter.}
    A meter token, such as \texttt{<meter\_4/4>}, establishes the time signature at the first beat and whenever it changes.
    A separate token gives the zero-based eighth-note position within the measure: in $4/4$, \texttt{<eighth\_pos\_0>} marks a downbeat and \texttt{<eighth\_pos\_6>} the fourth beat.
    The AR vocabulary supports numerators 1--32 and denominators 1, 2, 4, 8, 16, and 32; the Prober's evaluated meter inventory is narrower.

    \item \textbf{Relative subbeat position.}
    Each event group begins with a shift from the preceding group, measured in sixteenth notes.
    For example, \texttt{<subbeat\_shift\_2>} advances by an eighth note, or half a quarter-note beat.
    Attributes at the same onset share an event group and require no intervening shift; an initial zero shift is allowed.

    \item \textbf{Melody pitch and track.}
    Each melody onset combines a MIDI pitch in $0$--$127$ with a source track, as in \texttt{<pitch\_78\_track\_0>}.
    Track~0 denotes vocal melody and Track~1 instrumental melody.
    Full melody includes both tracks, whereas vocal-only transcription includes Track~0.

    \item \textbf{Note duration.}
    A pitch can be followed by a duration token indexing a fixed set of 24 templates in sixteenth-note units.
    For example, \texttt{<duration\_1>} represents two sixteenth notes (an eighth note), and \texttt{<duration\_3>} represents four (a quarter note).
    Omitting a duration token assigns the shortest template, one sixteenth note.
    Silent gaps are implicit in note onsets and durations, so no separate rest token is needed.

    \item \textbf{Chord.}
    A chord token encodes its root, quality, and, where applicable, inversion, for example \texttt{<chord\_full\_F\#:maj>}.
    The full vocabulary contains 361 labels, including a no-chord symbol; the major/minor alternative contains 25.
    The active chord is emitted initially, at chord changes, and again at downbeats to anchor each measure.

    \item \textbf{Key.}
    Tokens encode 24 tonic--mode combinations, such as \texttt{<key\_F\#:minor>} in Figure~\ref{fig:sheetsage2-ar}.
    The active key is emitted at the first beat and subsequently when it changes.

    \item \textbf{Musical structure.}
    Section tokens, such as \texttt{<structure\_chorus>}, select among 23 labels.
    The active section is emitted at the first beat of a window and at subsequent section changes.
\end{itemize}

\paragraph{From tokens to a complete score.}
Greedy decoding uses a grammar mask for field ordering and token types.
Recordings longer than 300 seconds are split into the fewest 300-second windows such that consecutive windows overlap by at least 100 seconds. Prior overlap events become the next prefix, with timestamps rebased to the next window.

\definecolor{arTime}{HTML}{596470}
\definecolor{arRhythm}{HTML}{8A5A10}
\definecolor{arStructure}{HTML}{80648C}
\definecolor{arKey}{HTML}{477047}
\definecolor{arChord}{HTML}{A65B28}
\definecolor{arMelody}{HTML}{365F88}
\definecolor{arPrefix}{HTML}{535860}
\clearpage
\subsection{Actual token sequences and score excerpts}
\label{app:ar-examples}
The following examples are actual AR predictions, not reference scores. Tokens retain their original values and order; line breaks are for display. Colors identify \textcolor{arTime}{time/shift}, \textcolor{arRhythm}{rhythm}, \textcolor{arStructure}{form}, \textcolor{arKey}{key}, \textcolor{arChord}{chord}, and \textcolor{arMelody}{melody/duration}. The corresponding score applies key-aware chord spelling (Appendix~\ref{app:spelling}); canonical token names remain unchanged.
\begin{table}[!htbp]
\centering
\caption{\textbf{Opening metadata and melody.} RWC-MDB-P-2001 No.~1, 0.02--7.17\,s. The task prefix and first four complete measures are shown. The first event establishes meter, section, key, chord, and melody source.}
\label{tab:ar-opening}
\begingroup
\setlength{\tabcolsep}{4pt}
\renewcommand{\arraystretch}{1.18}
\fontsize{8}{10}\selectfont
\begin{tabularx}{\textwidth}{@{}>{\raggedright\arraybackslash}p{0.13\textwidth}>{\raggedright\arraybackslash}X@{}}
\toprule
Context & Original serialized tokens \\
\midrule
Task prefix & \mbox{\textcolor{arPrefix}{\ttfamily <|sos|>}} \mbox{\textcolor{arPrefix}{\ttfamily <|timestamp|>}} \mbox{\textcolor{arPrefix}{\ttfamily <|downbeat\_meter|>}} \mbox{\textcolor{arPrefix}{\ttfamily <|structure|>}} \mbox{\textcolor{arPrefix}{\ttfamily <|key|>}} \mbox{\textcolor{arPrefix}{\ttfamily <|chord\_full|>}} \mbox{\textcolor{arPrefix}{\ttfamily <|melody\_full|>}} \mbox{\textcolor{arPrefix}{\ttfamily <|out|>}} \\
\addlinespace[2pt]
Measure 1\newline 0.02--1.83\,s & \mbox{\textcolor{arTime}{\ttfamily <subbeat\_shift\_0>}} \mbox{\textcolor{arTime}{\ttfamily <time\_0.02s>}} \mbox{\textcolor{arRhythm}{\bfseries \ttfamily <meter\_4/4>}} \mbox{\textcolor{arRhythm}{\ttfamily <eighth\_pos\_0>}} \mbox{\textcolor{arStructure}{\ttfamily <structure\_intro>}} \mbox{\textcolor{arKey}{\bfseries \ttfamily <key\_G\#:minor>}} \mbox{\textcolor{arChord}{\ttfamily <chord\_full\_G\#:min>}} \mbox{\textcolor{arMelody}{\ttfamily <pitch\_68\_track\_1>}} \mbox{\textcolor{arMelody}{\ttfamily <duration\_4>}} \mbox{\textcolor{arTime}{\ttfamily <subbeat\_shift\_4>}} \mbox{\textcolor{arTime}{\ttfamily <time\_0.50s>}} \mbox{\textcolor{arRhythm}{\ttfamily <eighth\_pos\_2>}} \mbox{\textcolor{arTime}{\ttfamily <subbeat\_shift\_2>}} \mbox{\textcolor{arMelody}{\ttfamily <pitch\_68\_track\_1>}} \mbox{\textcolor{arTime}{\ttfamily <subbeat\_shift\_1>}} \mbox{\textcolor{arMelody}{\ttfamily <pitch\_70\_track\_1>}} \mbox{\textcolor{arTime}{\ttfamily <subbeat\_shift\_1>}} \mbox{\textcolor{arTime}{\ttfamily <time\_0.94s>}} \mbox{\textcolor{arRhythm}{\ttfamily <eighth\_pos\_4>}} \mbox{\textcolor{arMelody}{\ttfamily <pitch\_71\_track\_1>}} \mbox{\textcolor{arMelody}{\ttfamily <duration\_2>}} \mbox{\textcolor{arTime}{\ttfamily <subbeat\_shift\_3>}} \mbox{\textcolor{arMelody}{\ttfamily <pitch\_70\_track\_1>}} \mbox{\textcolor{arMelody}{\ttfamily <duration\_2>}} \mbox{\textcolor{arTime}{\ttfamily <subbeat\_shift\_1>}} \mbox{\textcolor{arTime}{\ttfamily <time\_1.39s>}} \mbox{\textcolor{arRhythm}{\ttfamily <eighth\_pos\_6>}} \mbox{\textcolor{arTime}{\ttfamily <subbeat\_shift\_2>}} \mbox{\textcolor{arMelody}{\ttfamily <pitch\_68\_track\_1>}} \mbox{\textcolor{arMelody}{\ttfamily <duration\_1>}} \\
\addlinespace[2pt]
Measure 2\newline 1.83--3.61\,s & \mbox{\textcolor{arTime}{\ttfamily <subbeat\_shift\_2>}} \mbox{\textcolor{arTime}{\ttfamily <time\_1.83s>}} \mbox{\textcolor{arRhythm}{\ttfamily <eighth\_pos\_0>}} \mbox{\textcolor{arChord}{\ttfamily <chord\_full\_F\#:maj>}} \mbox{\textcolor{arMelody}{\ttfamily <pitch\_68\_track\_1>}} \mbox{\textcolor{arMelody}{\ttfamily <duration\_4>}} \mbox{\textcolor{arTime}{\ttfamily <subbeat\_shift\_4>}} \mbox{\textcolor{arTime}{\ttfamily <time\_2.27s>}} \mbox{\textcolor{arRhythm}{\ttfamily <eighth\_pos\_2>}} \mbox{\textcolor{arTime}{\ttfamily <subbeat\_shift\_2>}} \mbox{\textcolor{arMelody}{\ttfamily <pitch\_66\_track\_1>}} \mbox{\textcolor{arMelody}{\ttfamily <duration\_1>}} \mbox{\textcolor{arTime}{\ttfamily <subbeat\_shift\_2>}} \mbox{\textcolor{arTime}{\ttfamily <time\_2.71s>}} \mbox{\textcolor{arRhythm}{\ttfamily <eighth\_pos\_4>}} \mbox{\textcolor{arMelody}{\ttfamily <pitch\_66\_track\_1>}} \mbox{\textcolor{arMelody}{\ttfamily <duration\_5>}} \mbox{\textcolor{arTime}{\ttfamily <subbeat\_shift\_4>}} \mbox{\textcolor{arTime}{\ttfamily <time\_3.17s>}} \mbox{\textcolor{arRhythm}{\ttfamily <eighth\_pos\_6>}} \\
\addlinespace[2pt]
Measure 3\newline 3.61--5.39\,s & \mbox{\textcolor{arTime}{\ttfamily <subbeat\_shift\_4>}} \mbox{\textcolor{arTime}{\ttfamily <time\_3.61s>}} \mbox{\textcolor{arRhythm}{\ttfamily <eighth\_pos\_0>}} \mbox{\textcolor{arChord}{\ttfamily <chord\_full\_E:maj7>}} \mbox{\textcolor{arMelody}{\ttfamily <pitch\_68\_track\_1>}} \mbox{\textcolor{arMelody}{\ttfamily <duration\_4>}} \mbox{\textcolor{arTime}{\ttfamily <subbeat\_shift\_4>}} \mbox{\textcolor{arTime}{\ttfamily <time\_4.05s>}} \mbox{\textcolor{arRhythm}{\ttfamily <eighth\_pos\_2>}} \mbox{\textcolor{arTime}{\ttfamily <subbeat\_shift\_2>}} \mbox{\textcolor{arMelody}{\ttfamily <pitch\_68\_track\_1>}} \mbox{\textcolor{arTime}{\ttfamily <subbeat\_shift\_1>}} \mbox{\textcolor{arMelody}{\ttfamily <pitch\_70\_track\_1>}} \mbox{\textcolor{arTime}{\ttfamily <subbeat\_shift\_1>}} \mbox{\textcolor{arTime}{\ttfamily <time\_4.50s>}} \mbox{\textcolor{arRhythm}{\ttfamily <eighth\_pos\_4>}} \mbox{\textcolor{arMelody}{\ttfamily <pitch\_71\_track\_1>}} \mbox{\textcolor{arMelody}{\ttfamily <duration\_2>}} \mbox{\textcolor{arTime}{\ttfamily <subbeat\_shift\_3>}} \mbox{\textcolor{arMelody}{\ttfamily <pitch\_70\_track\_1>}} \mbox{\textcolor{arMelody}{\ttfamily <duration\_2>}} \mbox{\textcolor{arTime}{\ttfamily <subbeat\_shift\_1>}} \mbox{\textcolor{arTime}{\ttfamily <time\_4.94s>}} \mbox{\textcolor{arRhythm}{\ttfamily <eighth\_pos\_6>}} \mbox{\textcolor{arTime}{\ttfamily <subbeat\_shift\_2>}} \mbox{\textcolor{arMelody}{\ttfamily <pitch\_71\_track\_1>}} \mbox{\textcolor{arMelody}{\ttfamily <duration\_1>}} \\
\addlinespace[2pt]
Measure 4\newline 5.39--7.17\,s & \mbox{\textcolor{arTime}{\ttfamily <subbeat\_shift\_2>}} \mbox{\textcolor{arTime}{\ttfamily <time\_5.39s>}} \mbox{\textcolor{arRhythm}{\ttfamily <eighth\_pos\_0>}} \mbox{\textcolor{arChord}{\ttfamily <chord\_full\_F\#:maj>}} \mbox{\textcolor{arMelody}{\ttfamily <pitch\_75\_track\_1>}} \mbox{\textcolor{arMelody}{\ttfamily <duration\_4>}} \mbox{\textcolor{arTime}{\ttfamily <subbeat\_shift\_4>}} \mbox{\textcolor{arTime}{\ttfamily <time\_5.83s>}} \mbox{\textcolor{arRhythm}{\ttfamily <eighth\_pos\_2>}} \mbox{\textcolor{arTime}{\ttfamily <subbeat\_shift\_2>}} \mbox{\textcolor{arMelody}{\ttfamily <pitch\_73\_track\_1>}} \mbox{\textcolor{arMelody}{\ttfamily <duration\_1>}} \mbox{\textcolor{arTime}{\ttfamily <subbeat\_shift\_2>}} \mbox{\textcolor{arTime}{\ttfamily <time\_6.27s>}} \mbox{\textcolor{arRhythm}{\ttfamily <eighth\_pos\_4>}} \mbox{\textcolor{arMelody}{\ttfamily <pitch\_73\_track\_1>}} \mbox{\textcolor{arMelody}{\ttfamily <duration\_5>}} \mbox{\textcolor{arTime}{\ttfamily <subbeat\_shift\_4>}} \mbox{\textcolor{arTime}{\ttfamily <time\_6.71s>}} \mbox{\textcolor{arRhythm}{\ttfamily <eighth\_pos\_6>}} \\
\addlinespace[2pt]
\bottomrule
\end{tabularx}
\endgroup
\par\vspace{8pt}
\includegraphics[width=\textwidth]{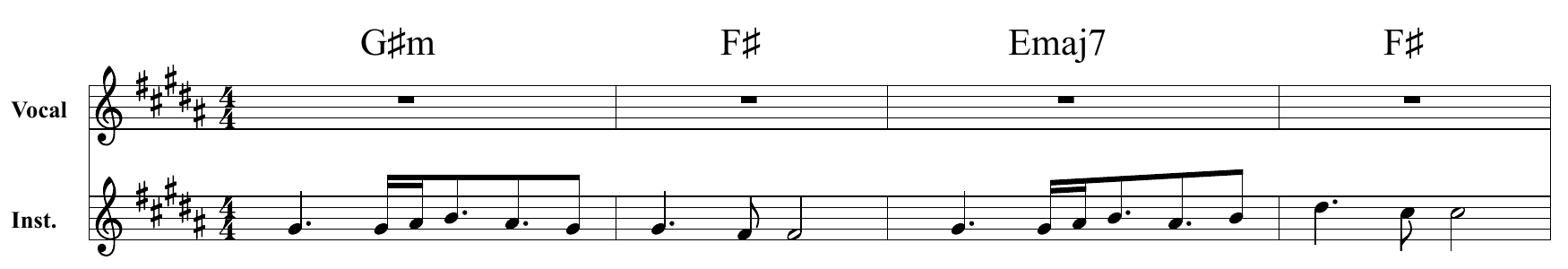}
\end{table}
\clearpage
\begin{table}[!htbp]
\centering
\caption{\textbf{A key change.} RWC-MDB-P-2001 No.~1, 35.61--39.17\,s. The key changes from G\(\sharp\) minor to C minor at 37.39\,s. The key token updates tonal context at the start of measure 22.}
\label{tab:ar-key-change}
\begingroup
\setlength{\tabcolsep}{4pt}
\renewcommand{\arraystretch}{1.18}
\fontsize{8}{10}\selectfont
\begin{tabularx}{\textwidth}{@{}>{\raggedright\arraybackslash}p{0.13\textwidth}>{\raggedright\arraybackslash}X@{}}
\toprule
Context & Original serialized tokens \\
\midrule
Measure 21\newline 35.61--37.39\,s & \mbox{\textcolor{arTime}{\ttfamily <subbeat\_shift\_2>}} \mbox{\textcolor{arTime}{\ttfamily <time\_35.61s>}} \mbox{\textcolor{arRhythm}{\ttfamily <eighth\_pos\_0>}} \mbox{\textcolor{arChord}{\ttfamily <chord\_full\_E:maj>}} \mbox{\textcolor{arMelody}{\ttfamily <pitch\_80\_track\_1>}} \mbox{\textcolor{arMelody}{\ttfamily <duration\_1>}} \mbox{\textcolor{arTime}{\ttfamily <subbeat\_shift\_2>}} \mbox{\textcolor{arMelody}{\ttfamily <pitch\_83\_track\_1>}} \mbox{\textcolor{arMelody}{\ttfamily <duration\_1>}} \mbox{\textcolor{arTime}{\ttfamily <subbeat\_shift\_2>}} \mbox{\textcolor{arTime}{\ttfamily <time\_36.05s>}} \mbox{\textcolor{arRhythm}{\ttfamily <eighth\_pos\_2>}} \mbox{\textcolor{arMelody}{\ttfamily <pitch\_82\_track\_1>}} \mbox{\textcolor{arMelody}{\ttfamily <duration\_1>}} \mbox{\textcolor{arTime}{\ttfamily <subbeat\_shift\_2>}} \mbox{\textcolor{arMelody}{\ttfamily <pitch\_83\_track\_1>}} \mbox{\textcolor{arTime}{\ttfamily <subbeat\_shift\_1>}} \mbox{\textcolor{arMelody}{\ttfamily <pitch\_85\_track\_1>}} \mbox{\textcolor{arMelody}{\ttfamily <duration\_3>}} \mbox{\textcolor{arTime}{\ttfamily <subbeat\_shift\_1>}} \mbox{\textcolor{arTime}{\ttfamily <time\_36.50s>}} \mbox{\textcolor{arRhythm}{\ttfamily <eighth\_pos\_4>}} \mbox{\textcolor{arChord}{\ttfamily <chord\_full\_F\#:maj>}} \mbox{\textcolor{arTime}{\ttfamily <subbeat\_shift\_4>}} \mbox{\textcolor{arTime}{\ttfamily <time\_36.94s>}} \mbox{\textcolor{arRhythm}{\ttfamily <eighth\_pos\_6>}} \mbox{\textcolor{arMelody}{\ttfamily <pitch\_87\_track\_1>}} \mbox{\textcolor{arMelody}{\ttfamily <duration\_3>}} \\
\addlinespace[2pt]
Measure 22\newline 37.39--39.17\,s & \mbox{\textcolor{arTime}{\ttfamily <subbeat\_shift\_4>}} \mbox{\textcolor{arTime}{\ttfamily <time\_37.39s>}} \mbox{\textcolor{arRhythm}{\ttfamily <eighth\_pos\_0>}} \mbox{\textcolor{arStructure}{\ttfamily <structure\_verse>}} \mbox{\textcolor{arKey}{\bfseries \ttfamily <key\_C:minor>}} \mbox{\textcolor{arChord}{\ttfamily <chord\_full\_C:min7>}} \mbox{\textcolor{arMelody}{\ttfamily <pitch\_70\_track\_1>}} \mbox{\textcolor{arMelody}{\ttfamily <duration\_1>}} \mbox{\textcolor{arTime}{\ttfamily <subbeat\_shift\_2>}} \mbox{\textcolor{arMelody}{\ttfamily <pitch\_70\_track\_0>}} \mbox{\textcolor{arMelody}{\ttfamily <duration\_1>}} \mbox{\textcolor{arTime}{\ttfamily <subbeat\_shift\_2>}} \mbox{\textcolor{arTime}{\ttfamily <time\_37.83s>}} \mbox{\textcolor{arRhythm}{\ttfamily <eighth\_pos\_2>}} \mbox{\textcolor{arMelody}{\ttfamily <pitch\_70\_track\_0>}} \mbox{\textcolor{arTime}{\ttfamily <subbeat\_shift\_1>}} \mbox{\textcolor{arMelody}{\ttfamily <pitch\_70\_track\_0>}} \mbox{\textcolor{arMelody}{\ttfamily <duration\_1>}} \mbox{\textcolor{arTime}{\ttfamily <subbeat\_shift\_2>}} \mbox{\textcolor{arMelody}{\ttfamily <pitch\_67\_track\_0>}} \mbox{\textcolor{arTime}{\ttfamily <subbeat\_shift\_1>}} \mbox{\textcolor{arTime}{\ttfamily <time\_38.27s>}} \mbox{\textcolor{arRhythm}{\ttfamily <eighth\_pos\_4>}} \mbox{\textcolor{arMelody}{\ttfamily <pitch\_70\_track\_0>}} \mbox{\textcolor{arTime}{\ttfamily <subbeat\_shift\_1>}} \mbox{\textcolor{arMelody}{\ttfamily <pitch\_70\_track\_0>}} \mbox{\textcolor{arMelody}{\ttfamily <duration\_1>}} \mbox{\textcolor{arTime}{\ttfamily <subbeat\_shift\_2>}} \mbox{\textcolor{arMelody}{\ttfamily <pitch\_70\_track\_0>}} \mbox{\textcolor{arMelody}{\ttfamily <duration\_1>}} \mbox{\textcolor{arTime}{\ttfamily <subbeat\_shift\_1>}} \mbox{\textcolor{arTime}{\ttfamily <time\_38.72s>}} \mbox{\textcolor{arRhythm}{\ttfamily <eighth\_pos\_6>}} \mbox{\textcolor{arTime}{\ttfamily <subbeat\_shift\_1>}} \mbox{\textcolor{arMelody}{\ttfamily <pitch\_67\_track\_0>}} \mbox{\textcolor{arTime}{\ttfamily <subbeat\_shift\_1>}} \mbox{\textcolor{arMelody}{\ttfamily <pitch\_67\_track\_0>}} \mbox{\textcolor{arTime}{\ttfamily <subbeat\_shift\_1>}} \mbox{\textcolor{arMelody}{\ttfamily <pitch\_66\_track\_0>}} \\
\addlinespace[2pt]
\bottomrule
\end{tabularx}
\endgroup
\par\vspace{8pt}
\includegraphics[width=\textwidth]{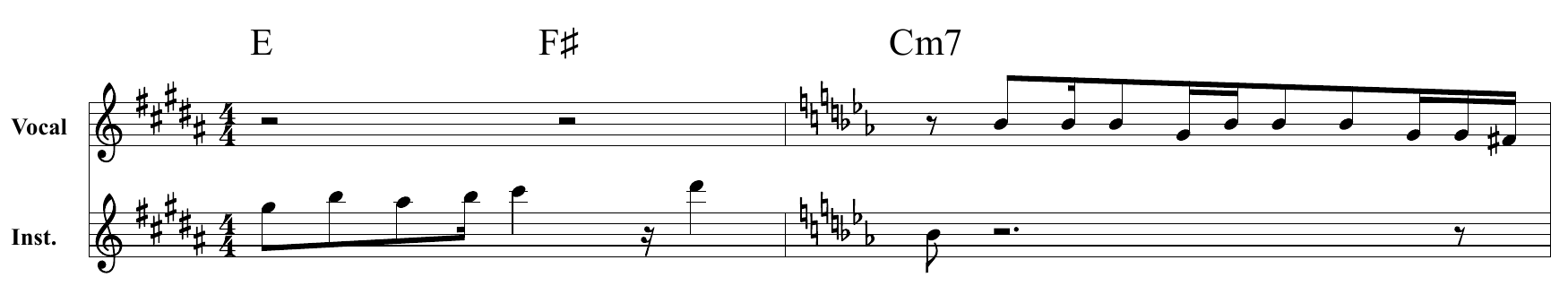}
\end{table}
\clearpage
\begin{table}[!htbp]
\centering
\caption{\textbf{A temporary meter change.} A song from Chords1217. The meter changes from $4/4$ to $3/4$ at 60.86\,s and returns to $4/4$ at 90.99\,s. The two excerpts show both boundaries; the 24 intervening measures are omitted.}
\label{tab:ar-meter-change}
\begingroup
\setlength{\tabcolsep}{4pt}
\renewcommand{\arraystretch}{1.18}
\fontsize{8}{10}\selectfont
\begin{tabularx}{\textwidth}{@{}>{\raggedright\arraybackslash}p{0.13\textwidth}>{\raggedright\arraybackslash}X@{}}
\toprule
Context & Original serialized tokens \\
\midrule
Measure 43\newline 59.40--60.86\,s & \mbox{\textcolor{arTime}{\ttfamily <subbeat\_shift\_4>}} \mbox{\textcolor{arTime}{\ttfamily <time\_59.40s>}} \mbox{\textcolor{arRhythm}{\ttfamily <eighth\_pos\_0>}} \mbox{\textcolor{arChord}{\ttfamily <chord\_full\_G\#:maj>}} \mbox{\textcolor{arMelody}{\ttfamily <pitch\_72\_track\_0>}} \mbox{\textcolor{arMelody}{\ttfamily <duration\_3>}} \mbox{\textcolor{arTime}{\ttfamily <subbeat\_shift\_4>}} \mbox{\textcolor{arTime}{\ttfamily <time\_59.76s>}} \mbox{\textcolor{arRhythm}{\ttfamily <eighth\_pos\_2>}} \mbox{\textcolor{arMelody}{\ttfamily <pitch\_72\_track\_0>}} \mbox{\textcolor{arMelody}{\ttfamily <duration\_1>}} \mbox{\textcolor{arTime}{\ttfamily <subbeat\_shift\_2>}} \mbox{\textcolor{arMelody}{\ttfamily <pitch\_73\_track\_0>}} \mbox{\textcolor{arMelody}{\ttfamily <duration\_1>}} \mbox{\textcolor{arTime}{\ttfamily <subbeat\_shift\_2>}} \mbox{\textcolor{arTime}{\ttfamily <time\_60.12s>}} \mbox{\textcolor{arRhythm}{\ttfamily <eighth\_pos\_4>}} \mbox{\textcolor{arChord}{\ttfamily <chord\_full\_A\#:min7>}} \mbox{\textcolor{arMelody}{\ttfamily <pitch\_73\_track\_0>}} \mbox{\textcolor{arMelody}{\ttfamily <duration\_3>}} \mbox{\textcolor{arTime}{\ttfamily <subbeat\_shift\_4>}} \mbox{\textcolor{arTime}{\ttfamily <time\_60.48s>}} \mbox{\textcolor{arRhythm}{\ttfamily <eighth\_pos\_6>}} \mbox{\textcolor{arMelody}{\ttfamily <pitch\_73\_track\_0>}} \mbox{\textcolor{arMelody}{\ttfamily <duration\_1>}} \mbox{\textcolor{arTime}{\ttfamily <subbeat\_shift\_2>}} \mbox{\textcolor{arMelody}{\ttfamily <pitch\_74\_track\_0>}} \mbox{\textcolor{arMelody}{\ttfamily <duration\_1>}} \\
\addlinespace[2pt]
Measure 44\newline 60.86--62.03\,s & \mbox{\textcolor{arTime}{\ttfamily <subbeat\_shift\_2>}} \mbox{\textcolor{arTime}{\ttfamily <time\_60.86s>}} \mbox{\textcolor{arRhythm}{\bfseries \ttfamily <meter\_3/4>}} \mbox{\textcolor{arRhythm}{\ttfamily <eighth\_pos\_0>}} \mbox{\textcolor{arStructure}{\ttfamily <structure\_bridge>}} \mbox{\textcolor{arKey}{\bfseries \ttfamily <key\_D:minor>}} \mbox{\textcolor{arChord}{\ttfamily <chord\_full\_G:min7>}} \mbox{\textcolor{arMelody}{\ttfamily <pitch\_77\_track\_0>}} \mbox{\textcolor{arMelody}{\ttfamily <duration\_3>}} \mbox{\textcolor{arTime}{\ttfamily <subbeat\_shift\_4>}} \mbox{\textcolor{arTime}{\ttfamily <time\_61.25s>}} \mbox{\textcolor{arRhythm}{\ttfamily <eighth\_pos\_2>}} \mbox{\textcolor{arMelody}{\ttfamily <pitch\_77\_track\_0>}} \mbox{\textcolor{arMelody}{\ttfamily <duration\_1>}} \mbox{\textcolor{arTime}{\ttfamily <subbeat\_shift\_2>}} \mbox{\textcolor{arMelody}{\ttfamily <pitch\_77\_track\_0>}} \mbox{\textcolor{arMelody}{\ttfamily <duration\_4>}} \mbox{\textcolor{arTime}{\ttfamily <subbeat\_shift\_2>}} \mbox{\textcolor{arTime}{\ttfamily <time\_61.65s>}} \mbox{\textcolor{arRhythm}{\ttfamily <eighth\_pos\_4>}} \\
\addlinespace[2pt]
\addlinespace
\multicolumn{2}{c}{\textit{24 intervening measures omitted}} \\
\addlinespace
Measure 69\newline 89.85--90.99\,s & \mbox{\textcolor{arTime}{\ttfamily <subbeat\_shift\_4>}} \mbox{\textcolor{arTime}{\ttfamily <time\_89.85s>}} \mbox{\textcolor{arRhythm}{\ttfamily <eighth\_pos\_0>}} \mbox{\textcolor{arChord}{\ttfamily <chord\_full\_G:maj>}} \mbox{\textcolor{arMelody}{\ttfamily <pitch\_77\_track\_0>}} \mbox{\textcolor{arMelody}{\ttfamily <duration\_3>}} \mbox{\textcolor{arTime}{\ttfamily <subbeat\_shift\_4>}} \mbox{\textcolor{arTime}{\ttfamily <time\_90.23s>}} \mbox{\textcolor{arRhythm}{\ttfamily <eighth\_pos\_2>}} \mbox{\textcolor{arMelody}{\ttfamily <pitch\_67\_track\_0>}} \mbox{\textcolor{arMelody}{\ttfamily <duration\_1>}} \mbox{\textcolor{arTime}{\ttfamily <subbeat\_shift\_2>}} \mbox{\textcolor{arMelody}{\ttfamily <pitch\_67\_track\_0>}} \mbox{\textcolor{arMelody}{\ttfamily <duration\_1>}} \mbox{\textcolor{arTime}{\ttfamily <subbeat\_shift\_2>}} \mbox{\textcolor{arTime}{\ttfamily <time\_90.61s>}} \mbox{\textcolor{arRhythm}{\ttfamily <eighth\_pos\_4>}} \mbox{\textcolor{arMelody}{\ttfamily <pitch\_67\_track\_0>}} \mbox{\textcolor{arMelody}{\ttfamily <duration\_3>}} \\
\addlinespace[2pt]
Measure 70\newline 90.99--92.45\,s & \mbox{\textcolor{arTime}{\ttfamily <subbeat\_shift\_4>}} \mbox{\textcolor{arTime}{\ttfamily <time\_90.99s>}} \mbox{\textcolor{arRhythm}{\bfseries \ttfamily <meter\_4/4>}} \mbox{\textcolor{arRhythm}{\ttfamily <eighth\_pos\_0>}} \mbox{\textcolor{arChord}{\ttfamily <chord\_full\_G:maj>}} \mbox{\textcolor{arMelody}{\ttfamily <pitch\_67\_track\_1>}} \mbox{\textcolor{arMelody}{\ttfamily <duration\_1>}} \mbox{\textcolor{arTime}{\ttfamily <subbeat\_shift\_2>}} \mbox{\textcolor{arMelody}{\ttfamily <pitch\_67\_track\_1>}} \mbox{\textcolor{arMelody}{\ttfamily <duration\_1>}} \mbox{\textcolor{arTime}{\ttfamily <subbeat\_shift\_2>}} \mbox{\textcolor{arTime}{\ttfamily <time\_91.35s>}} \mbox{\textcolor{arRhythm}{\ttfamily <eighth\_pos\_2>}} \mbox{\textcolor{arMelody}{\ttfamily <pitch\_67\_track\_1>}} \mbox{\textcolor{arMelody}{\ttfamily <duration\_1>}} \mbox{\textcolor{arTime}{\ttfamily <subbeat\_shift\_2>}} \mbox{\textcolor{arMelody}{\ttfamily <pitch\_67\_track\_1>}} \mbox{\textcolor{arMelody}{\ttfamily <duration\_1>}} \mbox{\textcolor{arTime}{\ttfamily <subbeat\_shift\_2>}} \mbox{\textcolor{arTime}{\ttfamily <time\_91.72s>}} \mbox{\textcolor{arRhythm}{\ttfamily <eighth\_pos\_4>}} \mbox{\textcolor{arMelody}{\ttfamily <pitch\_69\_track\_1>}} \mbox{\textcolor{arMelody}{\ttfamily <duration\_1>}} \mbox{\textcolor{arTime}{\ttfamily <subbeat\_shift\_2>}} \mbox{\textcolor{arMelody}{\ttfamily <pitch\_71\_track\_1>}} \mbox{\textcolor{arMelody}{\ttfamily <duration\_4>}} \mbox{\textcolor{arTime}{\ttfamily <subbeat\_shift\_2>}} \mbox{\textcolor{arTime}{\ttfamily <time\_92.09s>}} \mbox{\textcolor{arRhythm}{\ttfamily <eighth\_pos\_6>}} \\
\addlinespace[2pt]
\bottomrule
\end{tabularx}
\endgroup
\par\vspace{8pt}
{\footnotesize Measures 43--44 (59.40--62.03\,s)}\par
\includegraphics[width=\textwidth]{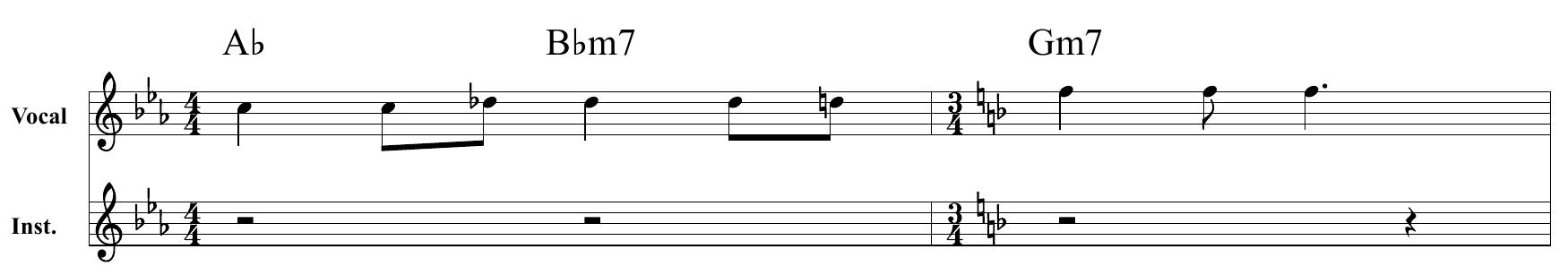}
{\footnotesize Measures 69--70 (89.85--92.45\,s)}\par
\includegraphics[width=\textwidth]{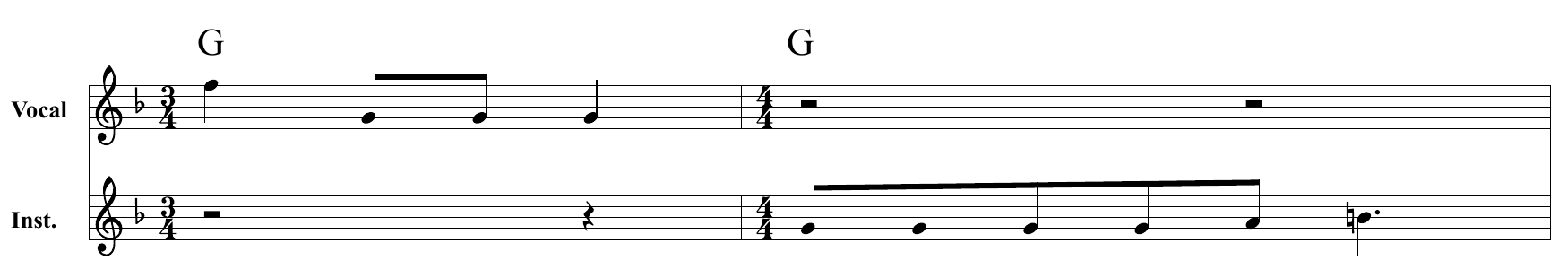}
\end{table}

\clearpage
\section{Training Data, Model Configurations, and Prober Outputs}
\label{app:training}

\subsection{Training sets}
The multitask Prober is trained on 393,980 songs totaling 22,016.7 hours of audio (Table~\ref{tab:source-pools}).
Most of this audio is rendered from MIDI; 32,755 songs are real recordings.
The Expanded lead-sheet corpus combines the Hooktheory annotations used by SheetSage~\citep{donahue2022sheetsage} and a private dataset of lead-sheet annotations, mostly of popular music.
The melody model is trained on 220,341 songs (12,560.9 hours) of rendered MIDI and real recordings with melody annotations.
SheetSage2-AR is trained on 441,094 recordings (28,434.3 hours) pseudo-labeled by Prober, consisting \versiontext{primarily of CC0 music and synthetic music.}{primarily of CC0 music and synthetic music; Tokenwave.AI provides most of the synthetic music under license.}

\begin{table}[htbp]
\centering\small
\caption{Training sets of the multitask Prober. Hours are audio durations; task labels may cover only part of a recording. LA: Los Angeles MIDI Dataset; MALD: MIDI AutoLabel Dataset, the task labels for LA; SLMS: Segmented Lakh MIDI Subset.}
\label{tab:source-pools}
\begin{tabular*}{\textwidth}{@{\extracolsep{\fill}}lrrl@{}}
\toprule
Training set & Songs & Audio hours & Audio type \\
\midrule
LA / MALD~\citep{lev2024losangelesmididataset,mald2025dataset} & 355,095 & 19,269.7 & MIDI synthesis \\
SLMS~\citep{eldeeb2025barwise} & 6,130 & 366.4 & MIDI synthesis \\
Expanded lead-sheet corpus~\citep{donahue2022sheetsage} & 32,243 & 2,349.1 & Recordings \\
HarmonixSet training set~\citep{nieto2019harmonix} & 512 & 31.5 & Recordings \\
\midrule
Total & 393,980 & 22,016.7 & Mixed \\
\bottomrule
\end{tabular*}
\end{table}

\subsection{Pseudo-label bootstrapping}
\label{app:pseudolabel-bootstrapping}
Pseudo-label bootstrapping aims to expand the availability of annotated MIDI files with automatic labelers. Each labeler fine-tunes the pretrained symbolic model of \citet{jiang2025versatile}, a hierarchical RoFormer~\citep{su2024roformer} trained autoregressively on unannotated Los Angeles MIDI files, with LoRA adapters and a small prediction head.
It learns from seed labels and is then applied to the whole Los Angeles MIDI collection.

\paragraph{Chord and key.}
We follow \citet{jiang2025musiclabelsitself}.
In a multitrack arrangement, some passages hint a chord/key label almost explicitly in a single track: sustained block chords reveal the harmony, a low sustained line reveals the bass, a seven-note diatonic note stream reveals the scale, and sustained final notes suggest the tonic.
Expert rules extract such local seed labels.
The track that supplies a label is then removed from the input, and the predictor learns to recover the label from the remaining tracks.
Because harmony and key are shared by the whole texture, the trained predictor can also label passages even if no track states them explicitly.
Chord predictions are decoded by template matching and key predictions by HMM smoothing.
For major/minor keys, the scale predictor is further fine-tuned on corrected RWC-Pop key labels~\citep{goto2002rwc,mald2025dataset}.

\paragraph{Melody.}
Melody differs from chord and key in two respects.
First, instead of content-based analysis, the seed labels come from track names: 17,134 MIDI files contain one or two non-drum tracks named ``melody'' (case-insensitive) with at least 50 notes in total.
Second, the predictor sees the complete arrangement, including the melody, and learns which notes form the melody at each sixteenth-note step.
At labeling time, its output selects a track rather than serving as the label.
We compare the predicted note onsets with each candidate track, choose the track with the highest onset F1, and regard it as the melody track only if this F1 is at least 0.7.
We then export all notes on the estimated melody track as the melody label.

\subsection{Models}
\label{app:model-configs}
\begin{figure}[ht]
\centering
\includegraphics[width=\textwidth]{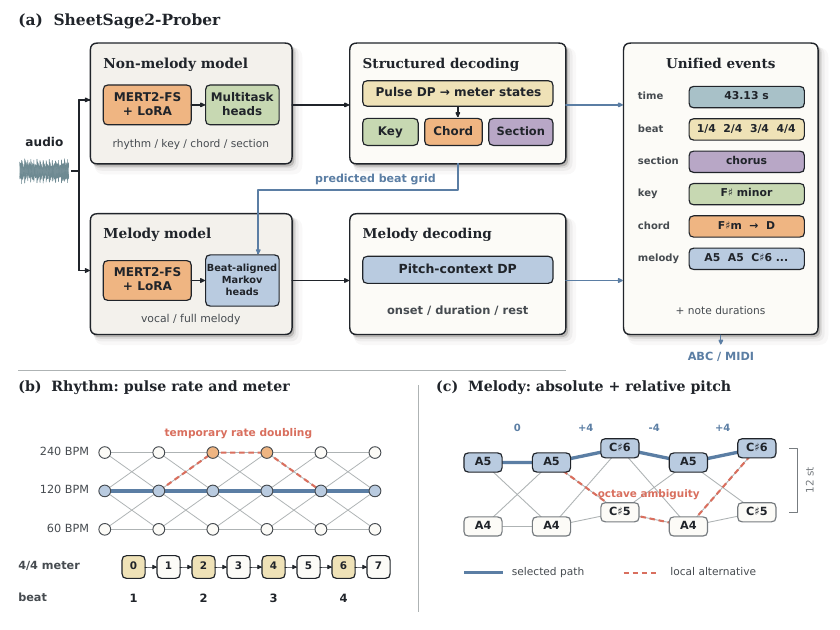}
\caption{SheetSage2-Prober and structured decoding.}
\label{fig:prober-architecture}
\end{figure}

Table~\ref{tab:model-configs} summarizes the three acoustic models, and Figure~\ref{fig:prober-architecture} shows how the two Prober models and their decoders are combined.
Each model adapts its own copy of the MERT2-FS encoder with LoRA (rank 64, $\alpha=128$) and is trained with AdamW in BF16 mixed precision.
Each evaluated checkpoint has the lowest validation loss of its training run.
At inference, all models run in BF16 without test-time augmentation.

\begin{table}[htbp]
\centering\small
\caption{Configurations of the evaluated models.}
\label{tab:model-configs}
\begin{tabularx}{\textwidth}{@{}l>{\raggedright\arraybackslash}X>{\raggedright\arraybackslash}X>{\raggedright\arraybackslash}X@{}}
\toprule
 & Non-melody Prober & Melody Prober & SheetSage2-AR \\
\midrule
Head & Two-layer MLP, width 512 & Six RoFormer layers, width 512, 8 heads & Six-layer BART decoder, width 512, 8 heads \\
Output & 160 binary channels & 926-class softmax & 31,678-class softmax over event tokens \\
Training data & 393,980 songs (Table~\ref{tab:source-pools}) & 220,341 songs & 441,094 pseudo-labeled recordings \\
Excerpts & 300~s and 30~s, 25-Hz frames & 30~s, 384 grid positions & 300~s, up to 5,120 tokens \\
Pitch augmentation & Continuous, $[-6,6)$ semitones & Continuous, $[-12,12)$ semitones & None \\
Global batch & 72 & 128 & 32 \\
Learning rate & $10^{-4}$, constant & $3\times10^{-4}$, constant & $3\times10^{-4}$, then cosine to $3\times10^{-5}$ \\
Optimizer steps & 57,500 & 72,500 & 207,000 + 6,750 \\
Inference window & 300~s (30~s for GTZAN) & 30~s, 15-s overlap & 300~s, overlap $\geq$100~s \\
\bottomrule
\end{tabularx}
\end{table}

\paragraph{Non-melody Prober.}
Each training recording contributes a 300-s excerpt and a 30-s excerpt with equal loss weight, so the model also sees inputs as short as GTZAN clips.
Pitch augmentation resamples all excerpts in a batch by one shared quasi-random shift in $[-6,6)$ semitones, which changes pitch and tempo together.
Pitch labels are transposed by the nearest semitone, and label times are rescaled.

\paragraph{Melody Prober.}
The melody model operates on subdivisions of the predicted beat grid and is conditioned on a vocal or full-melody hint; both hints share one model.
Pitch augmentation uses the same resampling, but each excerpt is resampled at two shifts drawn uniformly from $[-12,12)$ semitones; both shifts are shared by all excerpts in the batch.
Decoding uses the DP of Section~\ref{sec:melody-decoding} with temperature 1.
Before normalization, the rest logit is lowered by 1, which scales the rest probability by $e^{-1}\approx0.37$ relative to all other tokens, so the DP emits rests less often (Appendix~\ref{app:melody-dp}).

\paragraph{Melody Prober inputs.}
The melody model reads the encoder on the sixteenth-note grid: the MERT2-FS hidden states of all layers are projected to 512 dimensions and mixed with learned softmax weights, and the mixed frame at each grid position forms one input step.
Two learned 512-dimensional embeddings are added to every step before the six RoFormer layers (Table~\ref{tab:model-configs}): 
A beat embedding marks whether the step falls on a beat (value 1) or on a non-beat (value 0), and
a hint embedding, shared by all steps of an excerpt, identifies the melody annotation type of the training source: hint 0 for vocal-melody annotations, hint 1 for lead-melody annotations that may be instrumental, and hint 2 for synthetic LA melodies.
At inference, the same model produces the vocal output with hint 0 and the full-melody output with hint 1.

\paragraph{SheetSage2-AR.}
Training has two stages.
The first runs 207,000 steps at a constant learning rate.
The second starts from the best first-stage checkpoint with a fresh optimizer and anneals the learning rate over 6,750 steps.
Targets are the Prober's decoded event sequences with all timestamps shifted 30~ms earlier.
The AR model uses no pitch augmentation.

\subsection{Prober output channels}
\label{app:prober-outputs}
Table~\ref{tab:prober-channels} lists the Prober outputs.
Non-melody channels are independent binary outputs, whereas the melody outputs at each grid position share one softmax.

\begin{table}[htbp]
\centering\small
\caption{Prober output channels.}
\label{tab:prober-channels}
\begin{tabularx}{\textwidth}{@{}llr>{\raggedright\arraybackslash}X@{}}
\toprule
Model & Task & Channels & Content \\
\midrule
Non-melody & Chord & 48 & Root, chroma, extension, and bass, 12 pitch classes each \\
 & Key & 24 & 12 major and 12 minor keys \\
 & Rhythm & 64 & Downbeat, quarter-note, and eighth-note events, four 100-Hz slots each; two meter denominators; 25 median-tempo and 25 relative-tempo bins \\
 & Structure & 24 & 23 section classes and one boundary \\
\midrule
Melody & Melody & 926 & $128\times7$ pitch--interval categories, 24 durations, one rest, and five auxiliary slots \\
\bottomrule
\end{tabularx}
\end{table}

The chord decoder uses the root, chroma, and bass blocks.
Median-tempo bins have centers $\tau_\ell=30\cdot 2^{\ell/6}$ quarter-note BPM for $\ell=0,\ldots,24$, and relative-tempo bins have centers $2^{(\kappa-12)/6}$ for $\kappa=0,\ldots,24$ (Appendix~\ref{app:tempo-evidence}).
The median-tempo target is derived from the median eighth-note interval of the training excerpt, and relative-tempo targets divide the local tempo by this anchor.
The boundary target is positive within seven 25-Hz frames of a section start.
Appendix~\ref{app:melody-dp} gives the melody masks and decoder state.

\clearpage
\section{Exact Decoder Recurrences and Ablation Definitions}
\label{app:decoders}

This appendix describes the structured decoders used to convert frame-wise activations to discrete labels.
Indices are zero-based: $i$ indexes 100-Hz frames in rhythm decoding, sixteenth-note grid positions in melody decoding, and the original 25-Hz frames in key, section, and chord decoding; $j$ indexes recovered pulses or pooled segments, as specified below.
Each decoder maximizes its own score using dynamic programming.

\subsection{Pulse observations and tempo evidence}
\label{app:tempo-evidence}

The rhythm head provides 25-Hz frames with four fine event slots per frame.
Unfolding those slots produces a 100-Hz event sequence; dense denominator and tempo predictions are repeated four times.
Let $N_{\mathrm{raw}}$ be the length of this unfolded sequence, with downbeat, quarter-note, and eighth-note probabilities $d^{(i)}$, $b_4^{(i)}$, and $b_8^{(i)}$.

We use logarithmically spaced tempo bins. Global tempo bins are centered at $\tau_\ell=30\cdot 2^{\ell/6}$ quarter-note BPM for $\ell=0,\ldots,24$. Relative tempo bins are centered at $2^{(\kappa-12)/6}$ for $\kappa=0,\ldots,24$.
Let $g_\ell^{(i)}$ be the median-tempo prediction and $r_{\kappa}^{(i)}$ the relative-tempo prediction.
The decoder selects a single anchor from the neural predictions over the decoded input sequence:

\begin{equation}
\hat\ell=\arg\max_{\ell:\,60\leq\tau_\ell\leq240}\sum_{i=0}^{N_{\mathrm{raw}}-1}g_\ell^{(i)},
\qquad \hat g=\tau_{\hat\ell}.
\label{eq:tempo-anchor}
\end{equation}

Given anchor $\hat g$, relative bin $\kappa$ corresponds to local tempo $\hat g\cdot 2^{(\kappa-12)/6}$.
Let $v_\ell^{(i)}$ denote the resulting activations mapped onto the absolute tempo bins.
An outgoing pulse interval $\delta$ spans $\delta/100$ seconds; since two eighth-note pulses make one quarter note, the corresponding tempo is $\operatorname{BPM}(\delta)=60/(2\delta/100)=3000/\delta$.

Let $\ell(\delta)$ be the nearest bin to $\operatorname{BPM}(\delta)$ on the logarithmic tempo axis.
The compatibility score $W$ accumulates support for this tempo from the activations over the interval:

\begin{equation}
W\!\left(\hat g,\operatorname{BPM}(\delta);r^{(i:i+\delta)}\right)
=0.03\sum_{u=i}^{\min(i+\delta,N)-1}\max_{|\ell'-\ell(\delta)|\leq2}h(v_{\ell'}^{(u)}).
\label{eq:tempo-evidence}
\end{equation}

The maximum is over valid tempo bins, and $h$ is the observation transform defined in Section~\ref{sec:beat-decoding}.
Here $N$ is the retained sequence length after trimming trailing low-event-activation frames.
We evaluate Equation~\ref{eq:pulse-dp} for $i=0,\ldots,N-1$ and $\delta\in\mathcal D=\{12,\ldots,55\}$, corresponding to approximately 54.5--250 quarter-note BPM.
To let a path start, the predecessor maximum in Equation~\ref{eq:pulse-dp} also includes a start candidate $\sigma^{(i)}$, which is 0 for $0\leq i\leq\max\mathcal D$ and $-\infty$ otherwise; the maximum of an empty set is $-\infty$.
We select the highest-scoring pulse path that starts within 0.55 seconds of the input's beginning and whose final outgoing interval reaches or passes the last retained frame.

\subsection{Meter dynamic program}
\label{app:meter-dp}

The meter dynamic program assigns a position within the bar to each eighth-note pulse, identifying beats and downbeats while favoring a stable meter.
Let $t^{(j)}\in\{0,\ldots,N-1\}$ be the 100-Hz frame index of recovered pulse $j$, for $j=0,\ldots,J-1$.
Its meter state is $m^{(j)}=(n,\nu,Q,q)$, with $q\in\{0,\ldots,Q-1\}$.
The default $(n,\nu,Q)$ inventory is $(2,4,4)$, $(3,4,6)$, $(4,4,8)$, $(3,8,3)$, and $(6,8,6)$, giving 27 meter-position states.
Here $Q=8n/\nu$ is the nominal bar length in pulses, and $q=0$ denotes a downbeat.
An ordinary transition keeps $(n,\nu,Q)$ fixed and advances $q$ through $0\to1\to\cdots\to Q-1\to0$, maintaining complete bars.

For a transition from $m'=(n',\nu',Q',q')$ to $m=(n,\nu,Q,q)$, we allow two departures from this cycle:

\begin{enumerate}
\item \textbf{Local meter changes (LC)} accommodate incomplete bars and extra beats while retaining the nominal meter through early resets: $n=n'$, $\nu=\nu'$, and $q=0$ with $q'<Q'-1$.
For example, a bar containing only five eighth-note pulses within a $6/8$ passage follows $q=0\to1\to2\to3\to4\to0$, returning to the downbeat one pulse early while retaining the nominal $6/8$ meter.
Extra beats are represented by a complete bar followed by an additional shortened bar.

\item \textbf{Global meter changes (GC)} switch the nominal meter at a downbeat, such as $3/4\to4/4$: $n\ne n'$, $\nu=\nu'$, and $q=0$.
The denominator $\nu$ remains fixed throughout decoding.
\end{enumerate}

At its 100-Hz position $i=t^{(j)}$, the observation score is

\begin{equation}
U_{\mathrm{meter}}^{(j)}(m)=
\mathbf1[q=0]h(d^{(i)})
+\mathbf1[\nu=4\land q\equiv0\pmod2]h(b_4^{(i)})
+h(\eta_\nu^{(i)}).
\end{equation}

The observation score handles both denominator 4 and 8. For denominator 4, even positions $q$ are exported as beats; for denominator 8, every pulse is a beat.
In both cases, $q=0$ marks a downbeat. The transition score is

\begin{equation}
V_{\mathrm{meter}}(m',m)=
\begin{cases}
0,&\text{ordinary advance},\\
-8\bigl(1+I_{\mathrm{half}}(m')\bigr),&\text{local meter change (LC)},\\
-30-16I_{\mathrm{half}}(m'),&\text{global meter change (GC)},\\
-\infty,&\text{otherwise}.
\end{cases}
\label{eq:meter-transition}
\end{equation}

where $I_{\mathrm{half}}(m')=\mathbf1[\nu'=4\land (q'+1)\bmod2\ne0]$ indicates a reset halfway through a quarter-note beat.
Thus ordinary advances cost zero, local changes incur a smaller penalty than global changes, and half-beat resets receive an additional penalty.
Let $G^{(j)}(m)$ be the best score for the first $j+1$ pulses ending in state $m$.
The recurrence is

\begin{equation}
\begin{aligned}
G^{(0)}(m)&=U_{\mathrm{meter}}^{(0)}(m),\\
G^{(j)}(m)&=U_{\mathrm{meter}}^{(j)}(m)+\max_{m'}\left[G^{(j-1)}(m')+V_{\mathrm{meter}}(m',m)\right],\quad j\geq1.
\end{aligned}
\label{eq:meter-dp}
\end{equation}

\subsection{Melody dynamic program}
\label{app:melody-dp}

The melody decoder performs decoding on a sixteenth note grid, the same resolution as SheetSage1~\citep{donahue2022sheetsage}. Each decoded note is represented by a pitched onset, an optional duration token right in the next time step. If the duration token is absent, the note is decoded as a single sixteenth note. Possible duration token values (in sixteenth notes) are: 2, 3, 4, 6, 8, 12, 16, 24, 32, 48, 64, 96, 128, 192, 256, 384, 512, 768, 1024, 1536, 2048, 3072, and 4096.

Before the softmax, we mask the unused, start-of-sequence, mask, and padding tokens and the one-step duration, and lower the rest logit by 1; $\bar\mu^{(i)}$ denotes the resulting adjusted probabilities.
The DP state $(p,a)$ stores the last onset pitch and its age $a$ in grid positions, capped at 32.
With $\lambda_{\mathrm{rest}}^{(i)}$ and $\lambda_{\mathrm{dur}}^{(i)}$ the adjusted log-probabilities of a rest and of the best duration token (Table~\ref{tab:prober-channels}),
\begin{equation}
M^{(i)}(p,0)=\max_{p',a}\Bigl[M^{(i-1)}(p',a)+\log\bar\mu^{(i)}_{p,b}\Bigr],\qquad
b=\begin{cases}B(p-p'),&a<31,\\3,&a\geq31,\end{cases}
\end{equation}
\begin{equation}
M^{(i)}(p,a)=\max_{a':\,\min(a'+1,32)=a}M^{(i-1)}(p,a')+
\begin{cases}\max\bigl(\lambda_{\mathrm{rest}}^{(i)},\lambda_{\mathrm{dur}}^{(i)}\bigr),&a=1,\\ \lambda_{\mathrm{rest}}^{(i)},&a\geq2.\end{cases}
\end{equation}

Here, $a\geq 31$ falls back to the neutral bucket $b=3$ since the pitch transition over a long period is less meaningful.

The pitch-marginal ablation (Section~\ref{sec:octave-consistency}) replaces $\bar\mu^{(i)}_{p,b}$ in the first equation with $\sum_{b'}\bar\mu^{(i)}_{p,b'}$, so onset scores no longer depend on $p'$.

\subsection{DBN-only comparison}

The rhythm ablation supplies only the unfolded quarter-beat and downbeat probabilities to madmom's joint beat and downbeat DBN postprocessor~\citep{bock2016joint,madmom}.
It follows the Beat This! probability adaptation~\citep{foscarin2024beatthis}: with $\epsilon=10^{-5}$, set $\phi(v)=(1-\epsilon)v+\epsilon/2$ and calculate $[\max(\phi(b_4^{(i)})-\phi(d^{(i)}),\epsilon/2),\phi(d^{(i)})]$ as beat, downbeat activations, respectively.
The decoder uses 100 frames per second, beat counts per bar $\{3,4\}$, BPM limits 55 and 215, requested tempo-inventory size 60, transition parameter 100, observation parameter 16, threshold 0.05, and peak correction enabled.

\subsection{Key, section, and chord decoding}
\label{app:harmony-decoding}

These decoders operate on the previously decoded rhythm grid, and each maximizes its own sequence score.
Keys and sections are decoded per measure and chords per beat: predictions are pooled over each measure or beat, and a first-order dynamic program runs over the pooled sequence.
Here $i$ indexes the original 25-Hz predictions and $j$ indexes the pooled segments.

\paragraph{Keys.}
Keys are smoothed over measures by an HMM-style dynamic program with 24 states.
The observation score of each key is the logarithm of its component in the measure-averaged $k^{(i)}$, after adding $10^{-3}$.
Retaining the previous key costs zero and changing to another key costs 10, so a key change is kept only if it raises the path's summed observation score by more than 10.

\paragraph{Sections.}
Sections are also decoded per measure, with observation scores given by the logarithms of the measure-averaged $s^{(i)}$ plus $10^{-3}$.
Label changes alone cannot mark every boundary, because adjacent sections can share a label (e.g., two consecutive verses).
The decoder therefore also uses the boundary activation $s_o^{(i)}$.
For each downbeat, the boundary evidence $o^{(j)}$ is the maximum of $s_o^{(i)}$ between the midpoints of the preceding and current measures; the first window starts at the excerpt boundary.
Starting a new section at that downbeat costs $5(0.7-o^{(j)})$, and continuing the current section costs zero; the new section may keep the previous label.
When $o^{(j)}>0.7$, the cost becomes a reward and a new section always starts at that downbeat.
Below this threshold, a boundary requires a label change whose gain in summed observation score exceeds the cost.

\paragraph{Chords.}
Chords are decoded by template matching against a fixed vocabulary of 361 states: no chord (\texttt{N}) and 12 roots for each of the 30 quality--bass combinations in Table~\ref{tab:chord-vocabulary}.
Each chord state $z$ has a binary template $T_z\in\{0,1\}^{36}$ that concatenates a one-hot root, the quality's chroma template transposed to that root, and a one-hot bass; the template of \texttt{N} is zero.
The templates are compared with $\bar c^{(j)}$, the root, chroma, and bass components of the beat-averaged $c^{(i)}$; the extension head is unused.
With the all-ones vector $\mathbf1$, the observation score of state $z$ is
\begin{equation}
\label{eq:chord-template}
(\bar c^{(j)}-0.3\,\mathbf1)^\top(1.2\,T_z-0.4\,\mathbf1)
=1.2\,T_z^\top(\bar c^{(j)}-0.3\,\mathbf1)-0.4\,\mathbf1^\top(\bar c^{(j)}-0.3\,\mathbf1).
\end{equation}
The last term is shared by all states at the same beat and does not affect decoding.
Each active template entry therefore adds evidence in proportion to how far its predicted probability exceeds 0.3, and subtracts evidence when the probability is below 0.3.
The empty \texttt{N} template receives zero evidence, so apart from transition costs, \texttt{N} wins at a beat when every chord template's evidence is negative.
Changing chord costs 0.3; retaining it costs zero.
Adjacent identical chords are merged, then spelled using the key at each chord segment's midpoint (Appendix~\ref{app:spelling}).

\begin{table}[htbp]
\centering
\small
\caption{\textbf{Chord qualities in the decoder vocabulary.} Labels follow the Harte syntax~\citep{harte2005symbolic}. Filled cells mark each quality's chroma template, in semitones above the root. Every quality is used with all 12 roots in root position; the last column lists the additional bass options. As in the \texttt{mir\_eval} encoding~\citep{raffel2014mireval}, the bass note is always part of the chroma, so the \texttt{/2} states also contain the second; for example, \texttt{C:maj/2} has chroma C, D, E, G and bass D.}
\label{tab:chord-vocabulary}
\setlength{\tabcolsep}{3.2pt}
\begin{tabular}{@{}l*{12}{c}l@{}}
\toprule
Quality & 0 & 1 & 2 & 3 & 4 & 5 & 6 & 7 & 8 & 9 & 10 & 11 & Other bass \\
\midrule
\texttt{maj} & $\bullet$ &  &  &  & $\bullet$ &  &  & $\bullet$ &  &  &  &  & \texttt{/2}, \texttt{/3}, \texttt{/5} \\
\texttt{min} & $\bullet$ &  &  & $\bullet$ &  &  &  & $\bullet$ &  &  &  &  & \texttt{/2}, \texttt{/b3}, \texttt{/5} \\
\texttt{dim} & $\bullet$ &  &  & $\bullet$ &  &  & $\bullet$ &  &  &  &  &  & --- \\
\texttt{aug} & $\bullet$ &  &  &  & $\bullet$ &  &  &  & $\bullet$ &  &  &  & --- \\
\texttt{maj7} & $\bullet$ &  &  &  & $\bullet$ &  &  & $\bullet$ &  &  &  & $\bullet$ & \texttt{/3}, \texttt{/5}, \texttt{/7} \\
\texttt{min7} & $\bullet$ &  &  & $\bullet$ &  &  &  & $\bullet$ &  &  & $\bullet$ &  & \texttt{/b3}, \texttt{/5}, \texttt{/b7} \\
\texttt{7} & $\bullet$ &  &  &  & $\bullet$ &  &  & $\bullet$ &  &  & $\bullet$ &  & \texttt{/3}, \texttt{/5}, \texttt{/b7} \\
\texttt{hdim7} & $\bullet$ &  &  & $\bullet$ &  &  & $\bullet$ &  &  &  & $\bullet$ &  & --- \\
\texttt{dim7} & $\bullet$ &  &  & $\bullet$ &  &  & $\bullet$ &  &  & $\bullet$ &  &  & --- \\
\texttt{minmaj7} & $\bullet$ &  &  & $\bullet$ &  &  &  & $\bullet$ &  &  &  & $\bullet$ & --- \\
\texttt{sus2} & $\bullet$ &  & $\bullet$ &  &  &  &  & $\bullet$ &  &  &  &  & --- \\
\texttt{sus4} & $\bullet$ &  &  &  &  & $\bullet$ &  & $\bullet$ &  &  &  &  & --- \\
\texttt{sus4(b7)} & $\bullet$ &  &  &  &  & $\bullet$ &  & $\bullet$ &  &  & $\bullet$ &  & --- \\
\texttt{maj6} & $\bullet$ &  &  &  & $\bullet$ &  &  & $\bullet$ &  & $\bullet$ &  &  & --- \\
\texttt{min6} & $\bullet$ &  &  & $\bullet$ &  &  &  & $\bullet$ &  & $\bullet$ &  &  & --- \\
\bottomrule
\end{tabular}
\end{table}

\subsection{Spelling correction}
\label{app:spelling}

MIDI note numbers and pitch-class chord labels do not distinguish enharmonic spellings, but score notation must choose one.
Depending on the harmonic context, the same pitch may be written as E, F\(\flat\), or D\dsharp.
We therefore spell melody notes and chord roots with deterministic, key-dependent rules.
Each decoded key has a fixed name, which determines its key signature: the tonics are D\(\flat\), E\(\flat\), F\(\sharp\), A\(\flat\), and B\(\flat\) for the black-key major keys, and C\(\sharp\), D\(\sharp\), F\(\sharp\), G\(\sharp\), and B\(\flat\) for the black-key minor keys.

\paragraph{Melody.}
A fixed lookup table maps each MIDI pitch class to a note letter and accidental under the current key signature.
Table~\ref{tab:pitch-spelling} expresses this mapping as intervals from the tonic for major and minor keys.
Diatonic pitches follow the key signature; each chromatic pitch receives the spelling listed in the table.
For example, MIDI pitch 80 is written as G\(\sharp\)5 in C major: the rule selects an augmented fifth above C, rather than the enharmonically equivalent minor sixth A\(\flat\).
The lookup also permits double accidentals, such as F\dsharp{} for pitch class G in G\(\sharp\) minor.
When emitting ABC, the builder tracks accidentals by note letter within each measure and resets this state at barlines and key changes.

\begin{table}[htbp]
\centering
\caption{\textbf{Melody pitch spelling by key.} Semitone offsets are measured from the tonic modulo 12. Each entry gives the interval spelling selected by the fixed lookup; alternative enharmonic spellings are not selected by this rule.}
\label{tab:pitch-spelling}
\begin{tabular}{@{}rll@{}}
\toprule
Semitone offset & Major key & Minor key \\
\midrule
0 & Perfect unison & Perfect unison \\
1 & Augmented unison & Minor second \\
2 & Major second & Major second \\
3 & Augmented second & Minor third \\
4 & Major third & Major third \\
5 & Perfect fourth & Perfect fourth \\
6 & Augmented fourth & Augmented fourth \\
7 & Perfect fifth & Perfect fifth \\
8 & Augmented fifth & Minor sixth \\
9 & Major sixth & Major sixth \\
10 & Minor seventh & Minor seventh \\
11 & Major seventh & Major seventh \\
\bottomrule
\end{tabular}
\end{table}

\begin{figure}[htbp]
\centering
\includegraphics{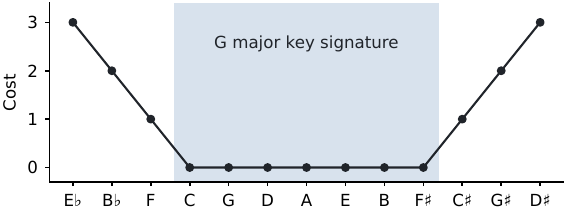}
\caption{\textbf{Tone cost for chord spelling in G major.} Notes are ordered on the line of fifths. The seven notes of the G major key signature (shaded) have zero cost, and the cost increases by one per fifth beyond either edge; it continues linearly outside the plotted range.}
\label{fig:spelling-cost}
\end{figure}

\paragraph{Chords.}
A chord label is spelled by choosing among candidate root spellings.
For each chord, we enumerate seven candidates, one for each letter C--B, with accidentals chosen to preserve the root pitch class; for pitch class C\(\sharp\), the candidates include C\(\sharp\), D\(\flat\), and B\dsharp.
Every chord tone is then spelled from the candidate root using the quality's interval template (e.g., 1, \(\flat\)3, \(\flat\)5, \(\flat\)7 for \texttt{hdim7}), so all tones keep their pitch classes.
Each tone is scored by its position on the line of fifths, i.e., the circle of fifths unrolled so that enharmonic spellings such as C\(\sharp\) and D\(\flat\) occupy different positions.
The seven notes of the key signature form a contiguous region with zero cost; outside it, the cost increases by one for each fifth beyond the nearest edge (Figure~\ref{fig:spelling-cost}).
Minor keys use the region of their relative major.
A candidate's score is the sum of its tone costs, with the root counted twice.
We select the lowest-scoring candidate and keep the chord quality and relative bass degree; ties follow the letter order C--B.
Table~\ref{tab:chord-spelling-example} shows the three lowest-scoring candidates for \texttt{C\#:hdim7} in G major.
The other four candidates need at least three accidentals on the root and score 96 or more.

\begin{table}[htbp]
\centering
\caption{\textbf{Spelling \texttt{C\#:hdim7} in G major.} Tone costs follow Figure~\ref{fig:spelling-cost}; the total counts the root cost twice. The selected spelling is C\(\sharp\):hdim7.}
\label{tab:chord-spelling-example}
\begin{tabular}{@{}lllr@{}}
\toprule
Spelling & Chord tones (1, \(\flat\)3, \(\flat\)5, \(\flat\)7) & Tone costs & Total \\
\midrule
C\(\sharp\):hdim7 & C\(\sharp\), E, G, B & 1, 0, 0, 0 & 2 \\
D\(\flat\):hdim7 & D\(\flat\), F\(\flat\), A\(\flat\flat\), C\(\flat\) & 5, 8, 11, 7 & 36 \\
B\dsharp:hdim7 & B\dsharp, D\dsharp, F\dsharp, A\dsharp & 13, 10, 7, 11 & 54 \\
\bottomrule
\end{tabular}
\end{table}

The Prober uses the decoded key at each chord segment's midpoint.
For the AR score excerpts in Appendix~\ref{app:ar-examples}, the same rule is applied to the chord timeline during notation export, using the key at each grid position.
For example, a \texttt{<chord\_full\_G\#:maj>} token under E\(\flat\) major is exported as \texttt{Ab:maj}.
All evaluation compares sounding pitches or pitch classes rather than written names, so these rules do not affect any reported score.

\clearpage
\section{Evaluation Protocols and Supplementary Benchmark Results}
\label{app:evaluation} 

\subsection{Evaluation datasets}
\label{app:eval-datasets}

\paragraph{Metrics.}
Beat and downbeat F1 discard reference and predicted events before 5~s, use a 70-ms tolerance, and are macro averaged over recordings.
Key evaluation uses the weighted score in Table~\ref{tab:key-score}, averaged over recordings.
For predictions that contain key changes, only the longest key segment is scored.
Chord scores average per-track major/minor recall weighted by reference annotation span.
Melody F1 is per-recording note F1 with a 50-ms onset tolerance and no offset requirement, macro averaged over recordings.
Octave errors (e.g., E4 vs E5) are ignored due to lack of reliable reference octaves.

\begin{table}[htbp]
\centering
\small
\caption{\textbf{Weighted key score.} Fifth errors in either direction receive 0.5.}
\label{tab:key-score}
\begin{tabular}{@{}llc@{}}
\toprule
Relation & Example & Score \\
\midrule
Same & C major vs.\ C major & 1.0 \\
Fifth & C major vs.\ G major & 0.5 \\
Relative & C major vs.\ A minor & 0.3 \\
Parallel & C major vs.\ C minor & 0.2 \\
Other & C major vs.\ D major & 0.0 \\
\bottomrule
\end{tabular}
\end{table}

\paragraph{GTZAN.}
We use 999 recordings for beat, 993 for downbeat, and 837 for key~\citep{tzanetakis2002musical}.
Beat and downbeat references use the GTZAN-Rhythm annotations~\citep{marchand2015swing}, matching those distributed with Beat This! annotations v1.0\footnote{\url{https://github.com/CPJKU/beat_this_annotations/tree/v1.0/gtzan}.}.
Key references come from \citet{kraft2013tonalness}, using v1.0 labels and excluding labels marked modulation/unknown\footnote{\url{https://github.com/audiocontentanalysis/dataset-gtzan-key/tree/v1.0}.}.

\paragraph{osu2017.}
Beat, downbeat, and chord evaluation use 142 recordings~\citep{jiang2026osu2017}.

\paragraph{GiantSteps.}
Key evaluation uses 604 recordings~\citep{knees2015two}.

\paragraph{Chords1217.}
Chord evaluation uses all 1,217 recordings~\citep{humphrey2015four}.

\paragraph{JAAH.}
Chord evaluation uses 113 recordings~\citep{eremenko2018jaah}.

\paragraph{HarmonixSet.}
Structure evaluation uses 200 test recordings~\citep{nieto2019harmonix}, a common 0.2-second frame grid, seven functional classes, and pooled one-to-one boundary matches at 0.5- and 3-second tolerances.

\paragraph{RWC-Pop.}
Vocal and full melody evaluation use all 100 recordings~\citep{goto2002rwc}.

\paragraph{Rock Corpus.}
Vocal melody evaluation uses all 200 recordings~\citep{temperley2013rock}.
Six of them have no pitched notes in their vocal references (rap or mostly unpitched vocals); they remain in the cohort and receive zero F1 for every system.
SheetSage1's melody output is scored against the vocal references; Prober and AR use their vocal outputs.
On the 194 recordings with nonempty references alone, SheetSage1, Prober, and AR obtain 50.71\%, 68.02\%, and 69.15\%, respectively.

\paragraph{Baselines.}
Beat This! is run under its original inference settings without DBN postprocessing, averaging the per-track scores of three released seeds.
SheetSage1 rhythm entries reuse madmom outputs and are not a separate rhythm inference experiment.
ChordFormer uses its released vocabulary and HMM decoder: a five-model ensemble on osu2017 and JAAH, and one held-out model per track on Chords1217.
This differs from applying a single fixed model to all Chords1217 tracks, and from the original scorer's pooling of metric-eligible durations.
The madmom chord model has training overlap with Chords1217.

\subsection{Chord metrics beyond major/minor}
\label{app:chord-metrics}
Table~\ref{tab:chord-metrics} reports the \texttt{mir\_eval} chord metrics used by ChordFormer~\citep{akram2025chordformer,raffel2014mireval}: root, thirds, maj/min, triads, sevenths, tetrads, and MIREX, and their inversion variants, which also require the correct bass.
Maj/min and sevenths ignore reference chords outside their vocabularies; sevenths, for example, covers maj, min, maj7, min7, 7, and no-chord.
AR scores highest on every metric on JAAH.
On osu2017 and Chords1217, Prober leads on root, thirds, maj/min, triads, and MIREX, AR on tetrads, and ChordFormer on sevenths, with and without inversions.

\begin{table}[htbp]
\centering\small
\caption{\textbf{Chord metrics (\%).} CF: ChordFormer, the strongest prior chord system in Table~\ref{tab:main-results}; on Chords1217 it uses five-fold cross-validation. Bold marks the best system for each benchmark and metric.}
\label{tab:chord-metrics}
\setlength{\tabcolsep}{3.5pt}
\begin{tabular}{@{}l*{9}{c}@{}}
\toprule
 & \multicolumn{3}{c}{osu2017} & \multicolumn{3}{c}{Chords1217} & \multicolumn{3}{c}{JAAH} \\
\cmidrule(lr){2-4}\cmidrule(lr){5-7}\cmidrule(lr){8-10}
Metric & CF & Prober & AR & CF & Prober & AR & CF & Prober & AR \\
\midrule
Root & 87.00 & \textbf{91.18} & 90.77 & 84.65 & \textbf{86.13} & 85.69 & 60.43 & 66.62 & \textbf{67.55} \\
Thirds & 85.72 & \textbf{90.21} & 89.91 & 81.79 & \textbf{83.04} & 82.55 & 57.56 & 63.02 & \textbf{64.14} \\
Maj/min & 86.55 & \textbf{90.43} & 90.08 & 83.94 & \textbf{84.29} & 83.81 & 59.45 & 62.94 & \textbf{64.50} \\
Triads & 83.96 & \textbf{88.26} & 87.99 & 77.73 & \textbf{78.80} & 78.53 & 56.65 & 60.27 & \textbf{61.81} \\
Sevenths & \textbf{76.19} & 74.64 & 75.58 & \textbf{72.39} & 71.60 & 72.07 & 46.49 & 45.15 & \textbf{47.89} \\
Tetrads & 72.61 & 72.08 & \textbf{73.04} & 65.82 & 65.94 & \textbf{66.60} & 40.04 & 40.22 & \textbf{43.00} \\
MIREX & 85.17 & \textbf{90.70} & 90.31 & 83.12 & \textbf{85.34} & 84.67 & 63.57 & 69.83 & \textbf{70.68} \\
\midrule
Thirds (inv.) & 83.09 & \textbf{86.99} & 86.44 & 79.73 & \textbf{81.01} & 80.72 & 55.86 & 60.04 & \textbf{61.80} \\
Maj/min (inv.) & 83.91 & \textbf{87.23} & 86.68 & 82.02 & \textbf{82.47} & 82.16 & 57.75 & 59.99 & \textbf{62.07} \\
Triads (inv.) & 81.44 & \textbf{85.15} & 84.67 & 76.01 & \textbf{77.20} & 77.07 & 55.07 & 57.50 & \textbf{59.56} \\
Sevenths (inv.) & \textbf{73.84} & 72.35 & 73.17 & \textbf{71.07} & 70.37 & 70.94 & 45.83 & 43.79 & \textbf{46.76} \\
Tetrads (inv.) & 70.38 & 69.87 & \textbf{70.70} & 64.67 & 64.86 & \textbf{65.63} & 39.53 & 38.83 & \textbf{41.94} \\
\bottomrule
\end{tabular}
\end{table}

\subsection{External Vocal-Melody Transcription Baselines}
\label{app:vocal-baselines}

Table~\ref{tab:main-results} lists SheetSage1 as the only prior melody system.
We also evaluate three recent note-level transcription systems with released weights, using the audio files, references, and metric of Table~\ref{tab:main-results}.
Each system follows its released inference settings, and we tune nothing on the test sets and apply no time shift or octave correction.

\begin{table}[h!]
\centering
\caption{\textbf{External vocal-melody transcription systems.} Benchmarks, metrics, and recordings match Table~\ref{tab:main-results}; external systems use their vocal output only. The Full row is grayed out because the external systems do not support melody reduction for non-vocal instruments. Scores are percentages; higher is better. Bold and underlining mark the largest and second-largest scores in each row.}
\label{tab:vocal-baselines}
\begingroup
\small
\setlength{\tabcolsep}{3pt}
\begin{tabular*}{\textwidth}{@{\extracolsep{\fill}}cccccccc@{}}
\toprule
\multirow{2}{*}{\textbf{Benchmark}} & \multirow{2}{*}{\textbf{Metric}} & \multirow{2}{*}{\textbf{SheetSage1}} & \multirow{2}{*}{\textbf{MuScriptor}} & \multirow{2}{*}{\textbf{YourMT3+}} & \multirow{2}{*}{\shortstack{\textbf{Demucs+}\\\textbf{ROSVOT}}} & \multicolumn{2}{c}{\textbf{SheetSage2}} \\
 & & & & & & \textbf{Prober} & \textbf{AR} \\
\midrule
\multirow{2}{*}{RWC-Pop} & Vocal F1\,$\uparrow$ & 62.71 & 47.04 & 49.12 & 26.09 & \textbf{83.08} & \underline{82.51} \\
 & \textcolor{gray}{Full F1\,$\uparrow$} & \textcolor{gray}{64.02} & \textcolor{gray}{38.71} & \textcolor{gray}{42.35} & \textcolor{gray}{22.30} & \textcolor{gray}{\underline{75.00}} & \textcolor{gray}{\textbf{75.29}} \\
Rock Corpus & Vocal F1\,$\uparrow$ & 49.19 & 37.36 & 36.05 & 19.86 & \underline{65.98} & \textbf{67.08} \\
\bottomrule
\end{tabular*}
\endgroup
\end{table}

\begin{itemize}
\item \textbf{MuScriptor}~\citep{rouard2026muscriptor}: we use the large model (1.4B parameters) to transcribe all instruments with greedy decoding, a classifier-free guidance coefficient of 1, and its default 5-s chunks. We score the notes of its voice class. Restricting generation to the voice class made it transcribe the accompaniment as voice, so we do not restrict it.
\item \textbf{YourMT3+}~\citep{chang2024yourmt3}: we use the released YPTF.MoE+Multi (noPS) checkpoint, the variant compared by \citet{rouard2026muscriptor}; we score its singing-voice (melody) class.
\item \textbf{Demucs+ROSVOT}: we first separate the vocals with HT Demucs~\citep{rouard2023hybrid} and then transcribe them with the ROSVOT transcriber~\citep{li2024rosvot}, using its default settings and predicted word boundaries.
Because ROSVOT is trained on utterances of at most about 21~s, we split the separated vocals at low-energy points into segments of at most 20~s.
\end{itemize}
TUTTI~\citep{hu2026tutti} is not included, as its weights are not released and it outputs untimed scores.

All three systems score far below SheetSage1 on every melody benchmark (Table~\ref{tab:vocal-baselines}).
The closest, YourMT3+, is below SheetSage1 by 13.59 points on RWC-Pop vocal and 13.14 points on Rock Corpus.

\paragraph{Global Time Offset.} Only ROSVOT gains much from shifting all of its notes by one constant offset chosen on the test sets (5-ms steps within $\pm$150~ms): delaying its notes by 45~ms on RWC-Pop and 50~ms on Rock Corpus raises its F1 to 35.29\%, 29.83\%, and 26.73\% in the three rows of Table~\ref{tab:vocal-baselines}, still far below SheetSage1.

\begin{figure}[h!]
\centering
\includegraphics[width=\textwidth]{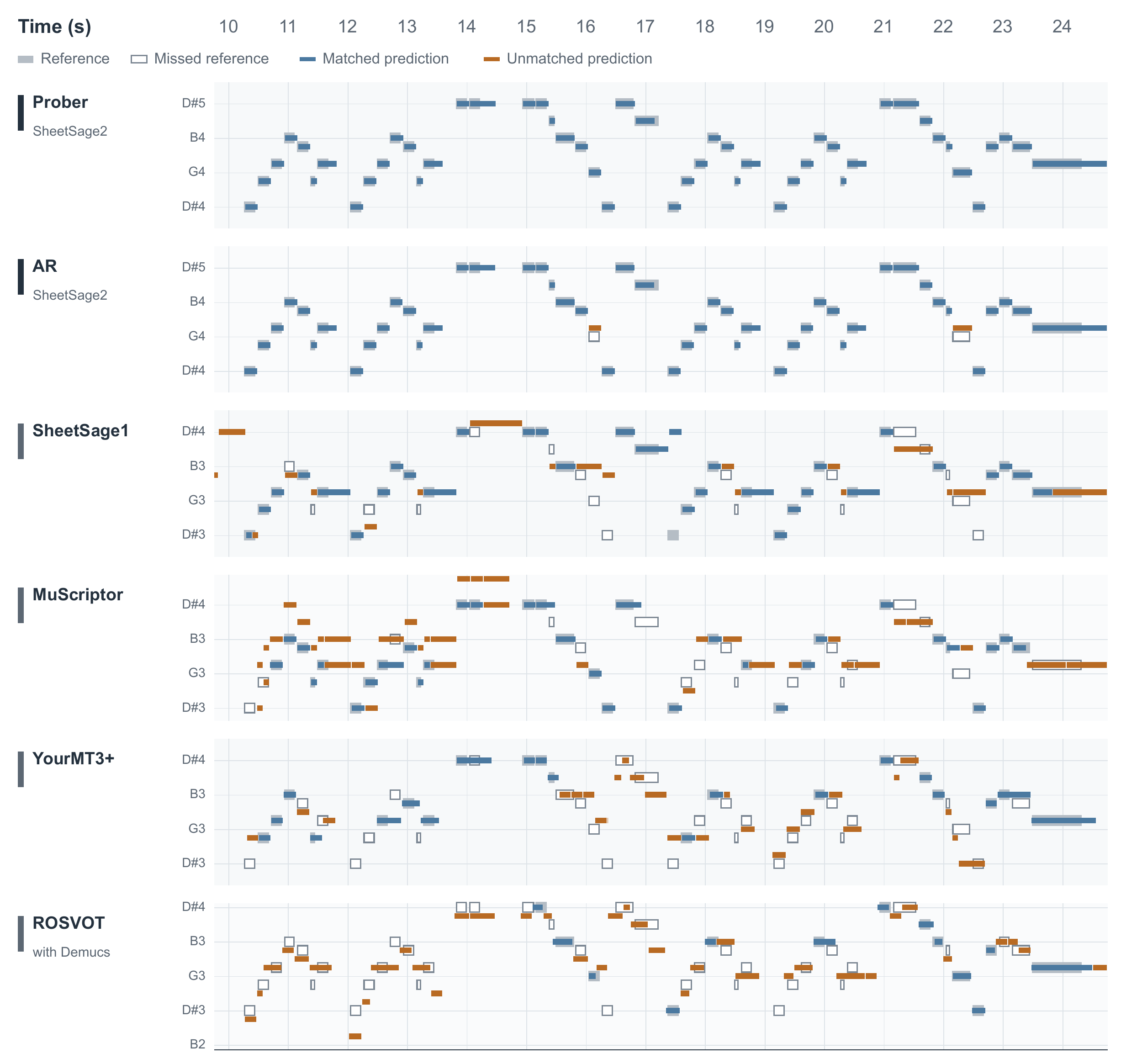}
\caption{\textbf{First sung phrase of RWC-MDB-P-2001 No.~1 (9.8--24.8~s).} Each panel shows one system of Table~\ref{tab:vocal-baselines}. Gray bars are reference notes, hollow where missed; blue and orange bars are predictions that do and do not match a reference note. The reference is shifted by whole octaves to best match each system.}
\label{fig:vocal-baselines-example}
\end{figure}

Figure~\ref{fig:vocal-baselines-example} shows the first sung phrase of RWC-MDB-P-2001 No.~1, the recording in Figure~\ref{fig:synth-melody-example}.
Prober matches every reference note. AR produces two pitch mismatches. SheetSage1 often merges repeated notes. MuScriptor hallucinates by adding a second voice. Many ROSVOT notes lie one semitone off the reference.

\clearpage
\subsection{Task-training data ablation}
\label{app:synthetic-data}
Table~\ref{tab:synthetic-data-full} expands Table~\ref{tab:synthetic-data} using exactly the tasks, benchmarks, metrics, and evaluation cohorts of Table~\ref{tab:main-results}.
Synthetic-only uses MALD~\citep{mald2025dataset} and SLMS~\citep{eldeeb2025barwise}, two labeled MIDI collections rendered to audio; real-only uses all other training datasets.

\begin{table}[htbp]
\centering
\caption{\textbf{Complete Prober data-source ablation.} Rows and metrics match Table~\ref{tab:main-results}; real + synthetic reproduces its Prober results. Scores are percentages; higher is better. Bold marks the largest point estimate in each row.}
\label{tab:synthetic-data-full}
\begingroup
\fontsize{10}{12}\selectfont
\setlength{\tabcolsep}{3pt}
\renewcommand{\arraystretch}{1.22}
\begin{tabular*}{\textwidth}{@{\extracolsep{\fill}}cccccc@{}}
\toprule
\multirow{2}{*}{\textbf{Task}} & \multirow{2}{*}{\textbf{Benchmark}} &
\multirow{2}{*}{\textbf{Metric}} & \multicolumn{3}{c}{\textbf{Task-training audio}} \\
 & & & \textbf{Real only} & \textbf{Synthetic only} & \textbf{Real + synthetic} \\
\midrule\midrule
\multirow{2}{*}{\textbf{Beat}} & GTZAN & \multirow{2}{*}{F1\,$\uparrow$} & 80.08 & 73.05 & \textbf{82.93} \\
 & osu2017 & & 90.79 & 85.98 & \textbf{92.28} \\
\cmidrule(lr){1-6}
\multirow{2}{*}{\textbf{Downbeat}} & GTZAN & \multirow{2}{*}{F1\,$\uparrow$} & 75.51 & 67.23 & \textbf{78.74} \\
 & osu2017 & & 88.60 & 86.05 & \textbf{92.79} \\
\cmidrule(lr){1-6}
\multirow{2}{*}{\textbf{Key}} & GiantSteps & \multirow{2}{*}{Score\,$\uparrow$} & 75.93 & 75.05 & \textbf{78.29} \\
 & GTZAN & & 75.46 & \textbf{76.26} & 72.62 \\
\cmidrule(lr){1-6}
\multirow{3}{*}{\textbf{Chord}} & osu2017 & \multirow{3}{*}{Maj/min\,$\uparrow$} & 89.69 & 88.65 & \textbf{90.43} \\
 & Chords1217 & & 81.95 & 82.04 & \textbf{84.29} \\
 & JAAH & & 61.32 & 60.28 & \textbf{62.94} \\
\cmidrule(lr){1-6}
\multirow{3}{*}{\textbf{Structure}} & \multirow{3}{*}{HarmonixSet} & Accuracy\,$\uparrow$ & 77.95 & 60.58 & \textbf{80.78} \\
 & & F1 (0.5\,s)\,$\uparrow$ & 66.32 & 49.55 & \textbf{66.95} \\
 & & F1 (3\,s)\,$\uparrow$ & 80.36 & 60.16 & \textbf{81.65} \\
\cmidrule(lr){1-6}
\multirow{3}{*}{\textbf{Melody}} & \multirow{2}{*}{RWC-Pop} & Vocal F1\,$\uparrow$ & 82.32 & 57.76 & \textbf{83.08} \\
 & & Full F1\,$\uparrow$ & 74.04 & 61.17 & \textbf{75.00} \\
 & Rock Corpus & Vocal F1\,$\uparrow$ & \textbf{67.67} & 45.37 & 65.98 \\
\bottomrule
\end{tabular*}
\endgroup
\end{table}

\begin{figure}[!t]
\centering
\includegraphics[width=\textwidth]{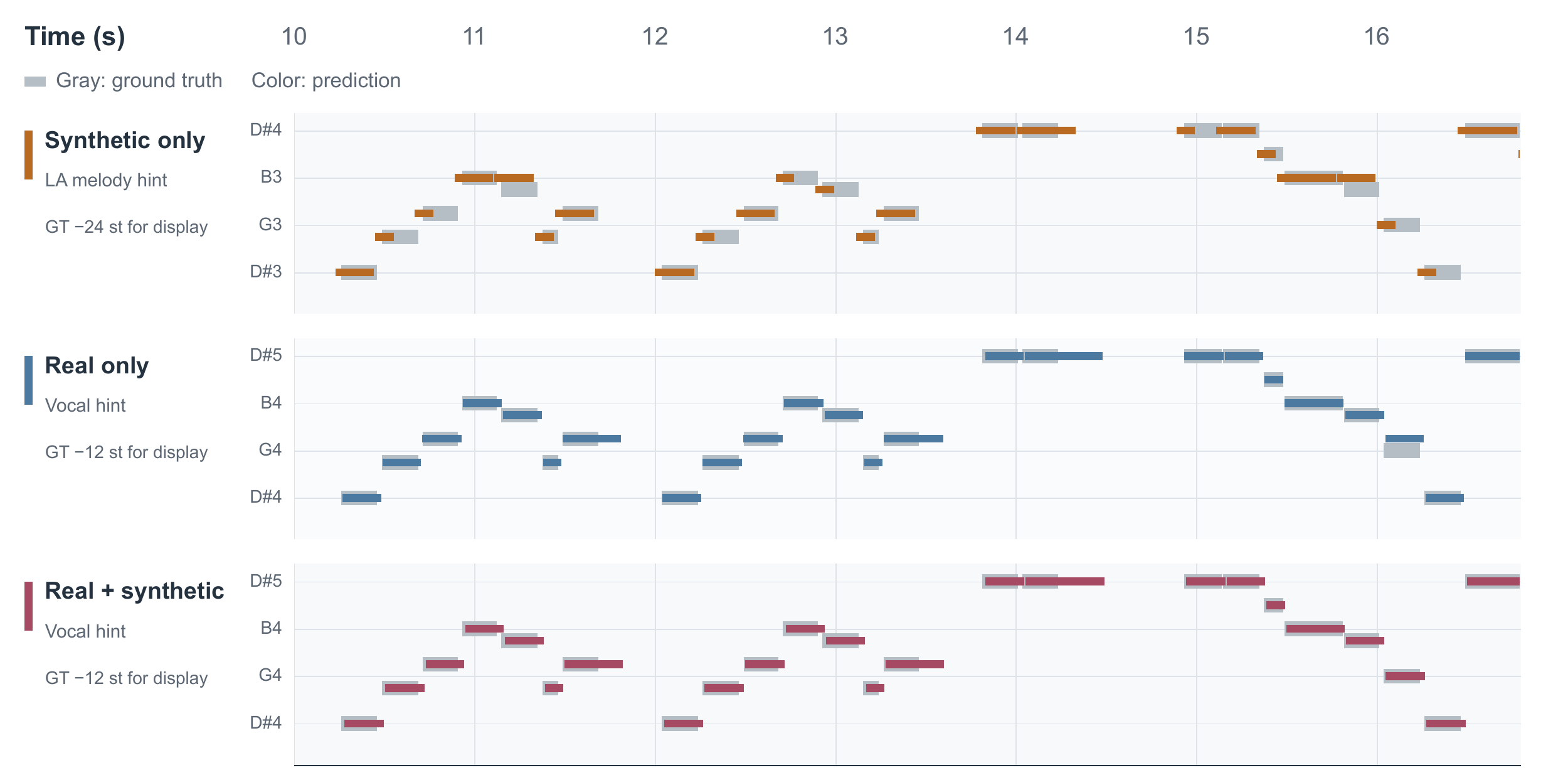}
\caption{\textbf{Opening vocal melody in RWC-MDB-P-2001 No.~1.} In each panel, colored Prober predictions overlay gray ground-truth notes over 10.0--16.8~s. All traces are the outputs scored in Table~\ref{tab:synthetic-data-full}. For comparison, the ground truth is shifted down one octave in the real-only and mixed panels and two octaves in the synthetic-only panel.}
\label{fig:synth-melody-example}
\end{figure}

Figure~\ref{fig:synth-melody-example} shows the first vocal phrase of RWC-MDB-P-2001 No.~1. The synthetic-only output follows much of the reference contour despite the absence of paired vocal audio in its task-training data. This selected example illustrates the behavior. The synthetic-only model still predicts vocal onsets and note lengths poorly.

\clearpage

\clearpage
\section{Consistency Diagnostics}
\label{app:consistency}

This appendix reports additional experiments on tempo and melody-octave consistency.

\subsection{Global Tempo Consistency}

We compare predicted and reference local tempos, which are given by inter-beat intervals.
Wherever a predicted interval overlaps a reference interval, the overlap is classified by the ratio of the predicted local tempo $\hat\tau$ to the reference local tempo $\tau$:
\begin{equation}
c=\begin{cases}
\text{half}, & \bigl|\log_2(\hat\tau/\tau)+1\bigr|\leq0.15,\\
\text{equal}, & \bigl|\log_2(\hat\tau/\tau)\bigr|\leq0.15,\\
\text{double}, & \bigl|\log_2(\hat\tau/\tau)-1\bigr|\leq0.15,\\
\text{other}, & \text{otherwise}.
\end{cases}
\end{equation}
Each class accumulates the overlap duration.
A recording is assessable when at least 80\% of its reference duration is classified as half, equal, or double.
It mixes tempo levels when at least two of these three classes each account for at least 5\% of the classified duration.

On GTZAN, osu2017, and Hooktheory, 888 of 999, 135 of 142, and 3,109 of 3,474 recordings are assessable for all four systems.
Unlike the 30-s GTZAN excerpts, osu2017 consists of full recordings; Hooktheory recordings are also transcribed in full but scored only on their annotated passages, at least two assessable passages per recording.
We exclude 166 Hooktheory recordings whose reference passages already differ in tempo by about a factor of two, leaving 2,943; without this exclusion, the ratios are 21.39\%, 5.63\%, 6.47\%, and 5.82\% for Beat This!, Prober, P-DBN, and AR.
Hooktheory overlaps the training data and is not a held-out test.
Beat This! uses its final0 checkpoint.

\begin{figure}[htbp]
\centering
\includegraphics[width=\textwidth]{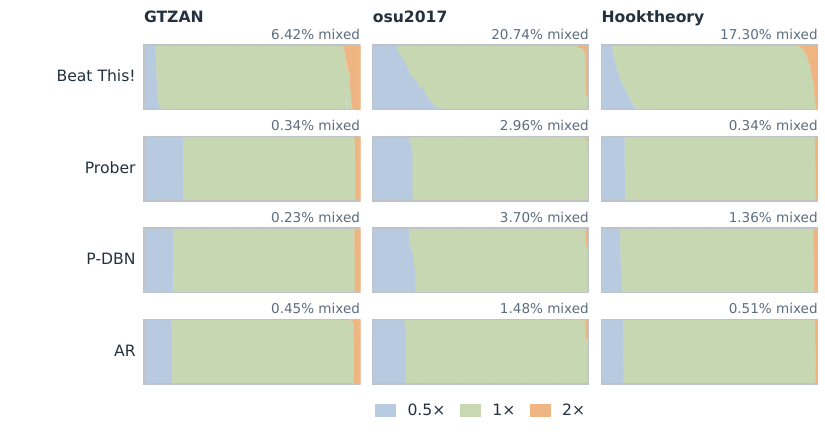}
\caption{\textbf{Tempo ratio composition.} Each vertical slice is one recording, sorted independently per strip by mean level; colors give the duration share at each tempo level relative to the reference. A recording mixes tempo levels when at least two colors each cover at least 5\% of its slice.}
\label{fig:consistency-tempo}
\end{figure}

Figure~\ref{fig:consistency-tempo} shows the distributions behind Table~\ref{tab:tempo-consistency}, together with GTZAN.
On osu2017 and Hooktheory, structured decoding mixes tempo levels less often than P-DBN, and AR distillation largely preserves this advantage.
GTZAN excerpts are too short to reveal long-range changes of tempo level; there, Prober and P-DBN mix levels in only 3 and 2 of 888 recordings, respectively.
On all three datasets, Prober and AR mix tempo levels far less often than Beat This!.

\subsection{Melody Octave Consistency}

For melody, we compare predicted and reference octaves on notes matched by onset (within 50~ms) and pitch class, recording the octave offset $\operatorname{round}((p_{\mathrm{pred}}-p_{\mathrm{ref}})/12)$ of each match.
Reference octaves are not always reliable, but a reference octave error usually shifts a whole passage uniformly.
We therefore measure consistency rather than accuracy: a constant offset within a unit is consistent, and a unit mixes octaves when at least two offsets each account for at least 5\% of its matched notes.

We compare SheetSage1, Prober, its pitch-marginal decoding, a direct-pitch Prober, and AR.
The direct-pitch Prober is trained from the same encoder and recipe but predicts 128 absolute pitches instead of the $128\times7$ pitch--interval categories, and is decoded like the pitch-marginal ablation; it thus tests the pitch--interval training targets rather than the decoder.

\begin{figure}[htbp]
\centering
\includegraphics[width=\textwidth]{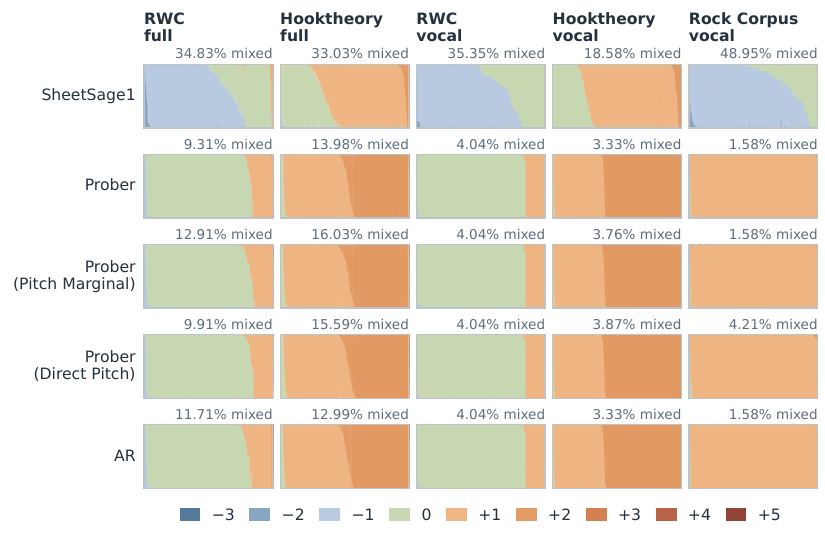}
\caption{\textbf{Octave-offset composition.} Each vertical slice is one unit (a passage or a recording), sorted independently per strip by mean offset; colors give the share of matched notes at each octave offset from the reference. A unit mixes octaves when at least two colors each cover at least 5\% of its slice.}
\label{fig:consistency-octave}
\end{figure}

On RWC full melody, a unit is a maximal contiguous run of one source instrument in the reference, and a later return to the same instrument starts a new passage.
RWC vocal and Rock Corpus vocal use whole recordings.
Hooktheory units are the annotated melody passages of 1,500 recordings drawn with a fixed seed from recordings with at least two passages; the three Prober variants share Prober's beat grid.
Every unit needs at least 20 note-level matches in every system (10 for Hooktheory); RWC passages and Rock Corpus recordings also need more than 40 reference notes, and Hooktheory passages at least 20.
This leaves 333 RWC full passages from 98 songs, 99 RWC vocal and 190 Rock Corpus recordings, and 2,925 full and 1,862 vocal Hooktheory passages; Figure~\ref{fig:consistency-octave} shows their offset distributions.
The Hooktheory recordings overlap the training data, so they do not form a held-out test.

The effect of pitch--interval decoding is clearer on full melody than on vocal melody.
On full melody, structured decoding (Prober) mixes octaves less often than both pitch-marginal decoding and the direct-pitch Prober.
AR mixes octaves less often than Prober on Hooktheory full but more often on RWC full.
On vocal melody, Prober improves little over pitch-marginal decoding; one possible reason is that vocal octaves are easier to model than those of other instruments.

SheetSage1 frequently switches octaves within a unit, the same kind of error that pitch marginalization introduces in Figure~\ref{fig:structured-decoding-examples}(c,d).
In every setting, all SheetSage2 variants mix octaves far less often than SheetSage1: SheetSage1 mixes octaves in 18.58--48.95\% of units, whereas no SheetSage2 variant exceeds 16.03\%.

\subsection{Transcription F1 of the Consistency Ablations}
\label{app:ablation-f1}

Table~\ref{tab:ablation-f1} reports F1 for the ablations above on the benchmarks, protocols, and recordings of Table~\ref{tab:main-results}.
We expect a trade-off between consistency and F1, especially for beats, but the results show no uniform trend: P-DBN scores higher than Prober on GTZAN and lower on osu2017.
For melody, Prober and pitch-marginal decoding of the same model reach nearly identical F1, differing by less than 0.1 points.
The direct-pitch Prober can improve Rock Corpus vocal F1 from 65.98\% to 69.54\%, but we favor consistency over a higher F1 on a single dataset.

\begin{table}[htbp]
\centering
\caption{\textbf{F1 of the consistency ablations.} Benchmarks, metrics, and recordings match Table~\ref{tab:main-results}, whose Prober scores are reproduced. Scores are percentages; higher is better. Bold marks the largest point estimate in each row; an em dash marks an ablation that does not apply.}
\label{tab:ablation-f1}
\begingroup
\small
\setlength{\tabcolsep}{3pt}
\begin{tabular*}{\textwidth}{@{\extracolsep{\fill}}ccccccc@{}}
\toprule
\textbf{Task} & \textbf{Benchmark} & \textbf{Metric} & \textbf{Prober} & \textbf{P-DBN} & \textbf{Pitch-marginal} & \textbf{Direct-pitch} \\
\midrule
\multirow{2}{*}{\textbf{Beat}} & GTZAN & \multirow{2}{*}{F1\,$\uparrow$} & 82.93 & \textbf{84.72} & --- & --- \\
 & osu2017 & & \textbf{92.28} & 91.55 & --- & --- \\
\cmidrule(lr){1-7}
\multirow{2}{*}{\textbf{Downbeat}} & GTZAN & \multirow{2}{*}{F1\,$\uparrow$} & 78.74 & \textbf{79.12} & --- & --- \\
 & osu2017 & & \textbf{92.79} & 90.65 & --- & --- \\
\cmidrule(lr){1-7}
\multirow{3}{*}{\textbf{Melody}} & \multirow{2}{*}{RWC-Pop} & Vocal F1\,$\uparrow$ & 83.08 & --- & 83.09 & \textbf{83.13} \\
 & & Full F1\,$\uparrow$ & 75.00 & --- & \textbf{75.02} & 74.85 \\
 & Rock Corpus & Vocal F1\,$\uparrow$ & 65.98 & --- & 66.07 & \textbf{69.54} \\
\bottomrule
\end{tabular*}
\endgroup
\end{table}

\clearpage
\definecolor{rwcKey}{HTML}{7B3FA0}
\definecolor{rwcMelody}{HTML}{D9A400}
\definecolor{rwcChord}{HTML}{2CA02C}
\definecolor{rwcSection}{HTML}{D62728}
\newcommand{\rwcbox}[2]{\textcolor{#1}{\textbf{#2}}}
\makeatletter\setlength{\@fptop}{0pt}\makeatother
\section{Lead-Sheet Excerpts from Full-Song Transcriptions}
\label{app:full-song}

This appendix shows excerpts of the lead sheets that SheetSage2-AR and SheetSage1 produce for two complete recordings, RWC-MDB-P-2001 No.~1 and No.~2 (Appendices~\ref{app:full-song-p001} and~\ref{app:full-song-p002}).
The recordings are from the RWC Music Database~\citep{goto2002rwc} and are released under CC BY-NC 4.0 (\url{https://creativecommons.org/licenses/by-nc/4.0/})~\citep{balke2026rwc}; the scores are automatic transcriptions of them, not their original scores.
SheetSage2-AR writes two staves, vocal and instrumental melody, together with chord symbols, key signatures, meter, tempo, and its predicted section labels (boxed).
SheetSage1 writes a single melody staff with chord symbols, under one key, meter, and tempo for the whole song.
The SheetSage2-AR score is rendered from its ABC export, and the SheetSage1 score is converted from its LilyPond output.
Both exporters split durations only at barlines, so for readability we re-spell both scores at beat boundaries (with ties) and beam by beat; pitches and sounding durations are unchanged.

Colored boxes mark notable disagreements with the RWC reference annotations of key, chords, melody, and section labels.
We selected them by hand and leave out minor errors.
A \rwcbox{rwcKey}{purple} dashed outline encloses a passage written in a key other than the reference key.
\rwcbox{rwcMelody}{Yellow} marks melody errors (wrong, missing, or extra notes, and wrong octaves), and \rwcbox{rwcChord}{green} marks chord-symbol errors.
\rwcbox{rwcSection}{Red} marks a section label that names the wrong section.

\begin{figure}[!b]
\centering
\includegraphics[width=\textwidth]{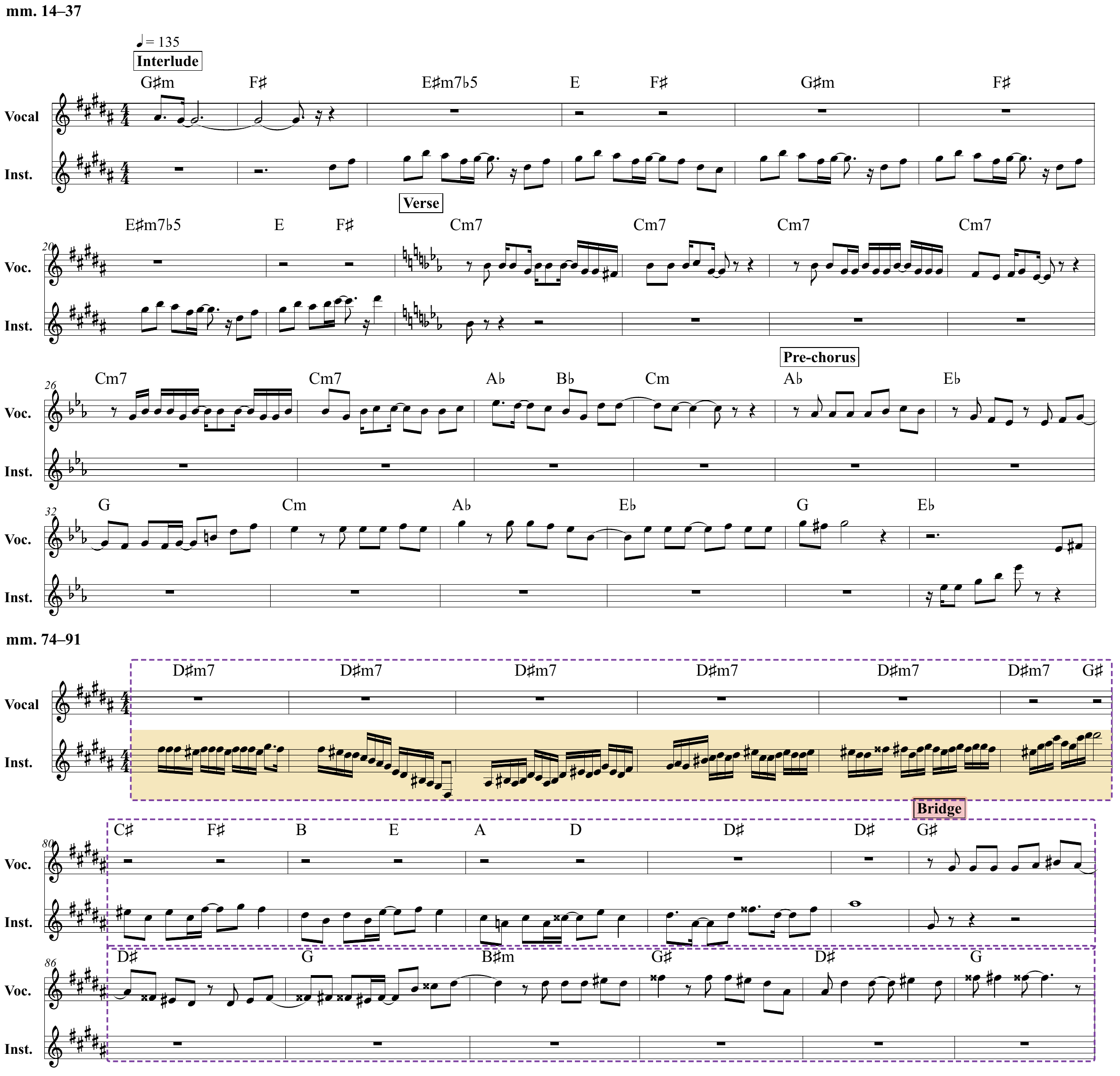}
\caption{SheetSage2-AR lead sheet for RWC-MDB-P-2001 No.~1, mm.~14--37 and 74--91 of 118.}
\label{fig:rwc-p001-ar}
\end{figure}

\subsection{RWC-MDB-P-2001 No. 1}
\label{app:full-song-p001}

Figures~\ref{fig:rwc-p001-ar} and~\ref{fig:rwc-p001-ss1} show mm.~14--37 and 74--91 of this recording, which is also used in Appendix~\ref{app:ar-examples}; the complete transcriptions have 118 measures (SheetSage2-AR) and 114 (SheetSage1).

\begin{figure}[!t]
\centering
\includegraphics[width=\textwidth]{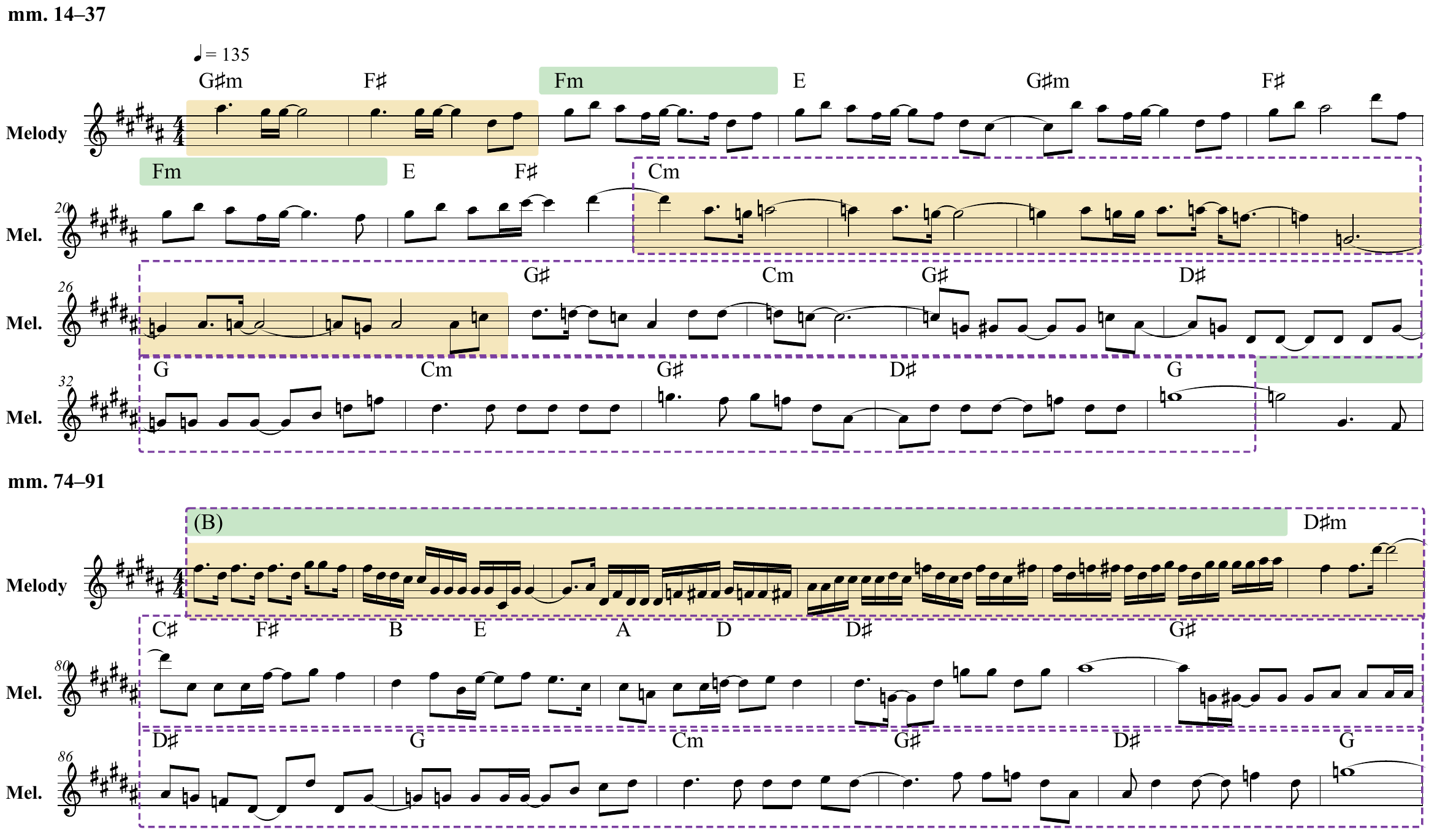}
\caption{SheetSage1 lead sheet for RWC-MDB-P-2001 No.~1, mm.~14--37 and 74--91 of 114.}
\label{fig:rwc-p001-ss1}
\end{figure}

\paragraph{Beat, Downbeat \& Meter} Both SheetSage1 and SheetSage2 establish the correct beat grid.

\paragraph{Key} SheetSage2-AR follows the modulation to C minor at the verse (m.~22) and returns to G$\sharp$ minor at the chorus just after the first excerpt (m.~38). However, from the guitar solo (m.~74) onward it stays in G$\sharp$ minor, missing the solo's modulation to E$\flat$ minor, the passing keys that follow (mm.~80--84), and the C minor verse (m.~85). SheetSage1, on the other hand, only estimates a global key, so it misses all of these key changes.

\paragraph{Chord} SheetSage2-AR's chord labels are generally acceptable. In mm.~16 and 20, E$\sharp$m7$\flat$5 is enharmonically equivalent to Fm7$\flat$5, the half-diminished seventh on the reference's F$^\circ$, and E$\sharp$m7$\flat$5 is the correct spelling in G$\sharp$ minor. SheetSage1 misses the diminished quality in these measures and writes Fm. Other chord issues are also marked in green.

\paragraph{Melody} SheetSage2-AR's melody transcription is generally acceptable, but the yellow passage (mm.~74--79) is a simplification of the original melody.
The original fast guitar solo has a subdivision beyond 16th notes, which is beyond SheetSage2-AR's capacity.
SheetSage1 contains more errors: it splits a single sung note into repeated onsets and misses the instrumental entry (mm.~14--15).
The rhythm of the rap is oversimplified (mm.~22--27).
The yellow passage (mm.~74--79) is further oversimplified compared to SheetSage2-AR.

\paragraph{Structure} SheetSage1 does not support structure label transcription. SheetSage2-AR starts each section at a downbeat, so a section that begins with a vocal pickup starts at the following bar line, which is a reasonable notational choice. After the solo, it labels the passage in C minor (m.~85) as a bridge (red), whereas the reference labels it a verse.

\begin{figure}[!t]
\centering
\includegraphics[width=\textwidth]{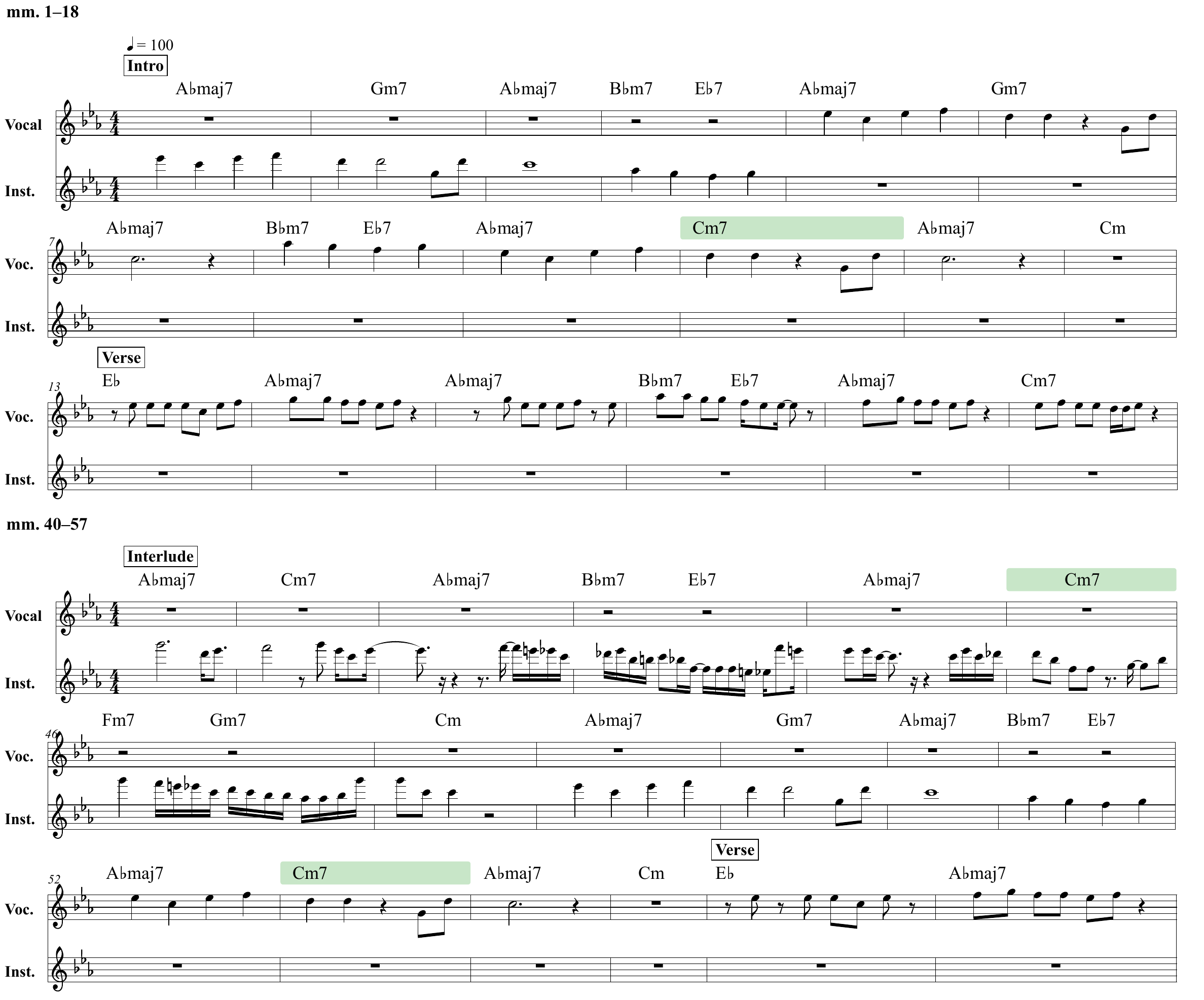}
\caption{SheetSage2-AR lead sheet for RWC-MDB-P-2001 No.~2, mm.~1--18 and 40--57 of 93.}
\label{fig:rwc-p002-ar}
\end{figure}

\subsection{RWC-MDB-P-2001 No.~2}
\label{app:full-song-p002}

\begin{figure}[!t]
\centering
\includegraphics[width=\textwidth]{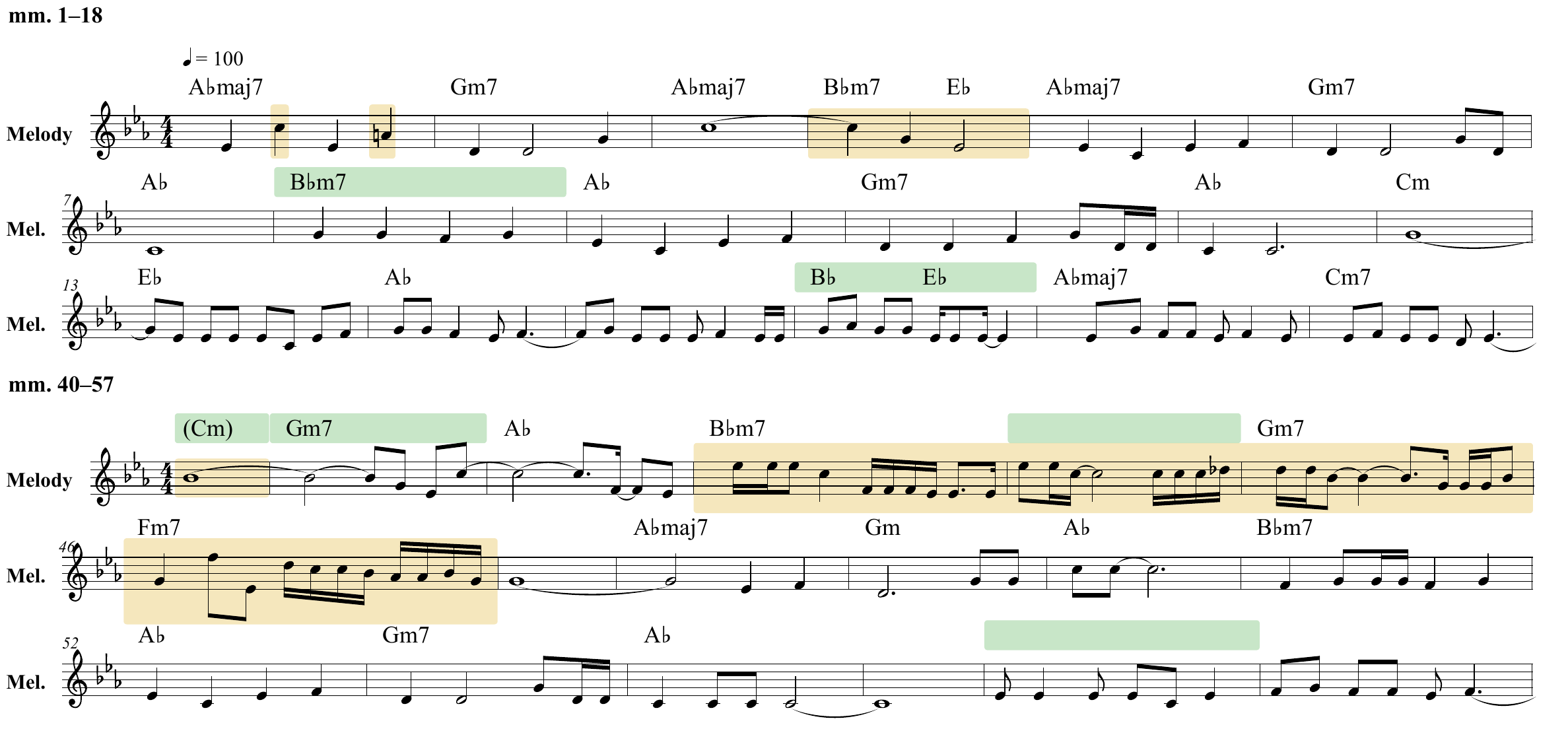}
\caption{SheetSage1 lead sheet for RWC-MDB-P-2001 No.~2, mm.~1--18 and 40--57 of 90.}
\label{fig:rwc-p002-ss1}
\end{figure}

Figures~\ref{fig:rwc-p002-ar} and~\ref{fig:rwc-p002-ss1} show mm.~1--18 (introduction, instrumental passage, and the start of the first verse) and mm.~40--57 (instrumental interlude and the start of the second verse) of this recording; the complete transcriptions have 93 measures (SheetSage2-AR) and 90 (SheetSage1).
The song is simpler than No.~1: 100 beats per minute in 4/4, with one key signature throughout.

\paragraph{Beat, Downbeat \& Meter} Both SheetSage1 and SheetSage2 establish the correct beat grid.

\paragraph{Key} Both systems choose E$\flat$ major, whereas the reference annotates its relative minor, C minor. Their key signatures are the same, and the chord progression strongly favors E$\flat$ major in most parts; only the cadence suggests C minor, so estimating E$\flat$ major is not a serious mistake.

\paragraph{Chord} SheetSage2-AR's chord labels are generally acceptable, but it writes Gm as Cm7 in mm.~10, 45, and 53, where SheetSage1 is correct. SheetSage1 writes wrong qualities (B$\flat$ for B$\flat$m7 in m.~16) or roots (B$\flat$m7 for E$\flat$7 in m.~8, Gm7 for Csus4 in m.~41), and it misses several chord changes (mm.~40, 44, and 56).

\paragraph{Melody} SheetSage2-AR transcribes the vocal melody of these excerpts almost exactly, and it follows the interlude melody, including its sixteenth-note runs, closely (mm.~40--50). We mark no melody error for SheetSage2-AR.
SheetSage1 again writes one staff without rests and contains more errors: a C an octave too high and an out-of-key A$\natural$ for F (m.~1) and a wrong motif (m.~4) in the introduction, a B$\flat$ held over the entry of the interlude melody (m.~40), and phrases that jump between octaves (mm.~43--46), for example G4--F5--E$\flat$4--D5 for the reference's G6--F6--E$\flat$6--D6 in m.~46.

\paragraph{Structure} SheetSage1 does not support structure label transcription. SheetSage2-AR's main error lies outside the excerpts: both choruses are labeled as verses. In this song the chorus is arranged much like the verse, but this is still a mistake. Other parts are largely reasonable: the verses and the interlude start at the reference boundaries (mm.~13, 40, and 56 in these excerpts), and merging the introduction with the following passage that repeats its material (mm.~1--12) is an acceptable choice.

\end{document}